\documentclass[11pt]{article} 
\usepackage[utf8]{inputenc} 
\usepackage[english]{babel}
\usepackage{geometry}
\usepackage{amsmath,amsfonts,amsthm,amssymb}
\usepackage{caption}
\usepackage{comment}
\usepackage{setspace}
\usepackage{geometry}
\usepackage{Tabbing}
\usepackage{lastpage}
\usepackage{tabularray}
\usepackage{extramarks}
\usepackage{chngpage}
\usepackage{systeme}
\usepackage{soul}
\usepackage{graphicx,float,wrapfig}
\usepackage[ruled,vlined,linesnumbered]{algorithm2e}
\usepackage{url}
\usepackage{dsfont}
\usepackage{multicol}
\usepackage{tikz}
\usepackage{tablefootnote}
\usetikzlibrary{trees,snakes}
\usetikzlibrary{arrows,automata}
\usetikzlibrary{matrix}
\usetikzlibrary{calc}

\usepackage{longtable} 

\usepackage{makecell} 
\usepackage{tabularx} 
\usepackage{booktabs,threeparttable} 
\usepackage{multirow} 

\usepackage{subcaption} 
\usepackage{mathtools}

\usepackage[ruled]{algorithm2e}
\usepackage{enumerate}
\usepackage{authblk}

\usepackage{natbib}
\usepackage{hyperref}
\usepackage{cleveref} 

\DeclareMathOperator{\vect}{vec}

\title{Dynamic Factor Analysis with Dependent Gaussian Processes
for High-Dimensional Gene Expression Trajectories}

\author[1]{Jiachen Cai, Robert J. B. Goudie, Colin Starr, Brian D. M. Tom}
\affil[1]{MRC Biostatistics Unit, University of Cambridge, UK}

\date{}

\begin{document}

\maketitle

\begin{abstract}
{The increasing availability of high-dimensional, longitudinal measures of gene expression can facilitate understanding of biological mechanisms, as required for precision medicine. Biological knowledge suggests that it may be best to describe complex diseases at the level of underlying pathways, which may interact with one another. We propose a Bayesian approach that allows for characterising such correlation among different pathways through Dependent Gaussian Processes (DGP) and mapping the observed high-dimensional gene expression trajectories into unobserved low-dimensional pathway expression trajectories via Bayesian Sparse Factor Analysis. Our proposal is the first attempt to relax the classical assumption of independent factors for longitudinal data and has demonstrated a superior performance in recovering the shape of pathway expression trajectories, revealing the relationships between genes and pathways, and predicting gene expressions (closer point estimates and narrower predictive intervals), as demonstrated through simulations and real data analysis. To fit the model, we propose a Monte Carlo Expectation Maximization (MCEM) scheme that can be implemented conveniently by combining a standard Markov Chain Monte Carlo sampler and an R package GPFDA \citep{konzen2021gaussian}, which returns the maximum likelihood estimates of DGP hyperparameters. The modular structure of MCEM makes it generalizable to other complex models involving the DGP model component. Our R package DGP4LCF that implements the proposed approach is available on CRAN.}

\textbf{Keywords}: {Dependent Gaussian processes; High-dimensional biomarker expression trajectories; Monte Carlo Expectation Maximization; Multivariate longitudinal data; Pathways; Sparse factor analysis.}
\end{abstract}

\section{Introduction}
The development of high-throughput technologies has enabled researchers to collect high-dimensional biomarker data repeatedly over time, such as transcriptome, metabolome and proteome, facilitating the discovery of disease mechanisms. Biological knowledge suggests that it may be best to describe complex diseases at the level of pathways, rather than the level of individual biomarkers \citep{richardson2016statistical}. A biological pathway is a series of interactions among molecules that achieves a certain biological function. Pathways are often summarized by activity scores derived from expression values of corresponding biomarker set \citep{temate2016inferring}; such scores form a basis of making further comparisons between people with different clinical statuses \citep{kim2021multi}.




This work was motivated by longitudinal measurements of gene expression in a human challenge study, where the main goal is to understand biological mechanism of influenza virus infection within the host, which provides a basis for developing novel diagnostic approaches for differentiating viral respiratory infection from respiratory disease caused by other pathogens \citep{zaas2009gene}. In brief, $17$ healthy individuals were inoculated with influenza virus H3N2, and then blood samples were collected at regular time intervals until the individuals were discharged after a fixed period of 7 days. The blood samples were assayed by DNA microarray, to produce gene expression values for $11,961$ genes. Binary labels denoting clinical status (symptomatic or asymptomatic) for each individual were also recorded. 

To handle such high-dimensional data, one commonly used dimension-reduction technique is factor analysis (FA). In the context of our motivating H3N2 example, FA maps the observed high-dimensional gene expression data to a latent low-dimensional pathway expression representation \citep{carvalho2008high}. One classical assumption of FA is the independence among latent factors, which implies expressions of pathways are uncorrelated. However, it has been shown that pathways often interact with one another to achieve complex biological functions \citep{hsu2012discovering}, so this assumption may not hold biologically. 

There have been attempts to relax this assumption of independence for cross-sectional data. Oblique rotation has been proposed as an alternative to standard orthogonal rotation under the frequentist framework \citep{costello2019best}, and \cite{conti2014bayesian} considered correlated factors within a Bayesian framework. However, for longitudinal data, no available method can characterize the potential cross-correlation between factor trajectories. To our knowledge, this paper is the first attempt to address this gap. Specifically, we model the latent factor trajectories using dependent Gaussian Processes (DGP), a tool from the machine learning community that allows for estimating cross-correlations and borrowing strength from correlated factors. Compared to an approach that assumes independence between factor trajectories, our approach performs better at recovering the shape of latent factor trajectories, estimating the relationship between genes and pathways, and predicting future gene expression. 

In addition to the modeling innovation, another contribution is the algorithm developed for estimating hyperparameters of the DGP model when it is embedded within another model. To obtain the maximum likelihood estimate (MLE) of DGP hyperparameters, we developed a Monte Carlo Expectation Maximization (MCEM) algorithm, which can be conveniently implemented by combining an existing R package, GPFDA \citep{konzen2021gaussian}, with a standard Markov Chain Monte Carlo (MCMC) sampler. Our R package DGP4LCF (Dependent Gaussian Processes for Longitudinal Correlated Factors) that implements the proposed method is available on CRAN.

The remainder of the article is organized as follows. In Section~\ref{Model}, we review Bayesian sparse factor analysis (BSFA) and DGP, then propose our integrated model based on them. In Section~\ref{mcem}, we introduce the inference method for the proposed model. We explore the empirical performance of our proposed approach for prediction of gene expression, estimation of gene-pathway relationships and the shape of pathway expression trajectories under several simulated factor generation mechanisms in Section~\ref{simulation_study}. In Section~\ref{data_application}, we compare our results on real data with those of a previous analysis in \cite{chen2011predicting}. We conclude with a discussion on future research directions in Section~\ref{discussion}.

\section{Model}
\label{Model}
Our proposed model is based on BSFA and DGP. Therefore, before presenting the proposed model in Section~\ref{proposed_model}, we first introduce BSFA in Section~\ref{uncover_underlying_structure} and DGP in Section~\ref{model_correlated_time_dependent_factor_trajectories}.

Let $t_{ij}$ denote the $j$th measured time point of the $i$th individual, $i=1,\dots,n$, $j=1,\dots,q_i$, where $n$ and $q_i$ are the number of subjects and subject-specific time points, respectively. At time $t_{ij}$, $x_{ijg}$ is the $g$th gene expression, $g=1,\dots,p$, where $p$ is the number of genes. We seek to describe these data in terms of $k$ latent factors/pathways $y_{ija}$, $a = 1,\dots,k$. In practice, we expect $k$ to be much smaller than $p$. Throughout this paper we assume that $k$ is pre-specified and fixed (see Section~\ref{choose_k} for discussion on the choice of $k$). 

\subsection{Uncovering sparse factor structure via BSFA}
\label{uncover_underlying_structure}
BSFA is an established approach to uncover a sparse factor structure. Sparsity stems from the fact that each pathway (i.e., factor) only involves a small proportion of genes (i.e., observed variable). Mathematically, BSFA connects observations $x_{ijg}$ with the unobserved latent factors $y_{ija}$ via a factor loading matrix $\mathbf{L}=\{l_{ga}\}_{g = 1,..., p, a = 1,...,k} \in \mathbb{R}^{p \times k}$, where each element $l_{ga}$ quantifies the extent to which the $g$th gene expression is related to the $a$th pathway expression, with larger absolute values indicating a stronger loading of the gene expression on the pathway expression.
\begin{equation}
   x_{ijg} = \mu_{ig} + \sum_{a=1}^{k}l_{ga}y_{ija} + e_{ijg},
\label{factor_analysis_scaler_representation}
\end{equation}
where $\mu_{ig}$ is the intercept term for the $g$th gene of the $i$th individual (hereafter the ``subject-gene mean''), $e_{ijg}$ is the residual error.
We assume ${l}_{ga}$ is constant across all time points.

We incorporate the prior belief of sparsity by imposing a sparsity-inducing prior distribution on $\mathbf{L}$. Specifically we adopt point-mass mixture priors \citep{carvalho2008high} because we want the model to shrink insignificant parameters completely to zero, without the need to further set up a threshold for inclusion, as in continuous shrinkage priors \citep{bhattacharya2011sparse}. The point-mass mixture prior is introduced by first decomposing $l_{ga}  = Z_{ga} \cdot A_{ga}$, where the binary variable $Z_{ga}$ indicates inclusion and the continuous variable $A_{ga}$ denotes the regression coefficient. We then specify a Bernoulli-Beta prior for $Z_{ga}$ with
${Z}_{ga} \sim \text{Bern}(\pi_a)$, 
$\pi_a \sim \text{Beta}(c_0, d_0)$ 
and a Normal-Inverse-Gamma prior for $A_{ga}$ with
${A}_{ga} \sim \text{N}(0,\rho_a^2)$, 
$\rho_a^2 \sim \text{Inverse-Gamma}(c_1, d_1)$,
where $g=1,\ldots,p; a=1,\ldots,k$ and $c_0, d_0, c_1, d_1$ are pre-specified positive constants. An {\it{a priori}} belief about sparsity can be represented via $(c_0, d_0)$, which controls the proportion of genes $\pi_a$ that contributes to the $a$th pathway. If $Z_{ga}=0$, meaning that the $g$th gene does not relate to the $a$th pathway, then the corresponding loading $l_{ga}=0$. 

We 
assign normal priors to the subject-gene means $\mu_{ig} \sim \text{N}(\mu_g, \sigma_g^2)$,
where $\mu_g$ is fixed to the mean of the $g$th gene expression across all time points of all people, and the residuals $e_{ijg} \sim \text{N}(0, \phi_g^2)$; inverse gamma priors to the variances $\sigma_g^2 \sim\text{Inverse-Gamma}(c_2, d_2)$
and $\phi_g^2 \sim\text{Inverse-Gamma}(c_3, d_3)$, where $c_2, d_2, c_3, d_3$ are pre-specified positive constants.

To implement the BSFA model, the software \href{https://www2.stat.duke.edu/~mw/mwsoftware/BFRM/index.html} {BFRM} has been developed \citep{carvalho2008high}. However, the current version of BFRM has two major limitations. First, it only returns point estimates of parameters, without any quantification of uncertainty. Second, it can handle only cross-sectional data; therefore it does not account for within-individual correlation when supplied with longitudinal data.

\subsection{Modeling correlated, time-dependent factor trajectories via DGP}
\label{model_correlated_time_dependent_factor_trajectories}

Several approaches treating factor trajectories ${y}_{a}(t)$, $a = 1,...,k$ as functional data have been proposed, including spline functions \citep{ramsay2009introduction}, differential equations \citep{james2019dynamic}, autoregressive models \citep{penny2002bayesian}, and Gaussian Processes (GP)\citep{shi2011gaussian}. As mentioned previously, we are interested in incorporating the cross-correlation among different factors into the model. GP are well-suited to this task because DGP can account for the inter-dependence of factors in a straight-forward manner. Indeed, the DGP model has been widely applied to model dependent multi-output time series in the machine learning community \citep{alvarez2011computationally}, where it is also known as ``multitask learning''. Sharing information between tasks using DGP can improve prediction compared to using independent Gaussian Processes (IGP) \citep{bonilla2007multi, cheng2020sparse}. DGP have also been used to model correlated, multivariate spatial data in the field of geostatistics, improving prediction performance over IGP \citep{dey2020graphical}. 

The main difficulty with DGP modeling is how to appropriately define cross-covariance functions that imply a positive definite covariance matrix. \cite{liu2018remarks} reviewed existing strategies developed to address this issue and found, through simulation studies, that no single approach outperformed all others in all scenarios. When the intent was to improve the predictions of
all the outputs jointly (such is our case here), the kernel convolution framework (KCF)\citep{boyle2005dependent} was among the best performers. The KCF has also been widely employed by other researchers \citep{alvarez2011computationally,shi2017regression}. Therefore, we adopt the KCF strategy for DGP modeling here, which assumes that the correlation between factors is the same at each time point. A detailed description of the KCF can be found in Supplementary A.2.

The distribution of $(y_{i1a}, \dots, y_{iq_ia}, y_{i1b},\dots, y_{iq_ib})^{T}$ induced under KCF is a multivariate normal distribution (MVN) with mean vector $\mathbf{0}$ and covariance matrix fully determined by parameters of kernel functions and a noise parameter; we use $\boldsymbol{\Theta}$ to denote all of them hereafter. The covariance matrix contains the information of both the auto-correlation for each single process and the cross-correlation between different processes. In this paper, we focus on the cross-correlation, which correspond to interactions across different biological pathways that have been ignored in previous analysis \citep{chen2011predicting}.

The KCF can be implemented via the R package GPFDA, which outputs the MLE for DGP hyperparameters given measurements of the processes ${y}_{a}(t), a = 1,\dots,k$. Its availability inspired and facilitated the algorithm developed for the proposed model. GPFDA assumes all input processes are measured at common time points, which is often unrealistic in practice; thus, in Section~\ref{mcem}, we will adapt GPFDA to subject-specific time points.

\subsection{Proposed integrated model}
\label{proposed_model}
We propose a model that combines BSFA and DGP (referred as ``BSFA-DGP'' hereafter), and present it in matrix notation below. To accommodate irregularly measured time points across individuals, we first introduce the vector of all unique observation times $\mathbf{t} = \bigcup\limits_{i=1}^{n}\mathbf{t}_{i}$ across all individuals, and denote its length as $q$; each $\mathbf{t}_{i}=\bigcup\limits_{j=1}^{q_i}t_{ij}$ is a vector of observed time points for the $i$th individual. We assume the irregularity of measurements is sufficiently limited such that Bayesian imputation is appropriate. Let $\mathbf{Y}_i = (\mathbf{y}_{i1},...,\mathbf{y}_{ik})^{T} \in \mathbb{R}^{k\times q}$ be the matrix of pathway expression of the $i$th individual, with $\mathbf{y}_{ia}= (y_{i1a}, ..., y_{iqa})^{T}$ denoting the $a$th factor's expression across all observation times $\mathbf{t}$; and let $\mathbf{Y}_{i,\text{obs}}, \mathbf{Y}_{i,\text{miss}}$ be the sub-matrices of $\mathbf{Y}_i$, denoting pathway expression at times when gene expression of the $i$th individual are observed and missing, respectively. Let $\vect({\mathbf{Y}_{i}}^{T})$ denote the column vector obtained by stacking the columns of matrix $\mathbf{Y}_i^{T}$ on top of one another; similar definitions apply to $\vect({\mathbf{Y}_{i,\text{obs}}^{T}})$ and $\vect({\mathbf{Y}_{i,\text{miss}}^{T}})$.

Let $\mathbf{X}_i=(\mathbf{x}_{i1},...,\mathbf{x}_{ip})^{T} \in \mathbb{R}^{p \times q_i}$ be the matrix of gene expression measurements at the $q_i$ observation times for the $i$th individual, with $\mathbf{x}_{ig}= (x_{i1g}, ..., x_{iq_{i}g})^{T}$ denoting the $g$th gene's trajectory; and correspondingly let $\mathbf{M}_{i} = (\boldsymbol{\mu}_{i1}, ..., \boldsymbol{\mu}_{ip})^{T} \in \mathbb{R}^{p \times q_i}$ be the matrix of subject-gene means, with $\boldsymbol{\mu}_{ig}=\mu_{ig}\mathbf{1}$, where $\mathbf{1}$ is a $q_i$-dimensional column vector consisting of the scalar $1$. Furthermore, let $\mathbf{A}=\{A_{ga}\}_{g=1,...,p; a = 1,...,k} \in \mathbb{R}^{p\times k}$ be the matrix of regression coefficients and $\mathbf{Z}=\{Z_{ga}\}_{g=1,...,p; a = 1,...,k} \in \mathbb{R}^{p\times k}$ be the matrix of inclusion indicators,
\begin{equation}
\begin{split}
        \mathbf{X}_i&=\mathbf{M}_{i} + \mathbf{L}\mathbf{Y}_{i, \text{obs}}+\mathbf{E}_{i},\\
        \mathbf{L}&=\mathbf{A}\circ \mathbf{Z},\\
        \vect({\mathbf{Y}_{i}}^{T}) & \sim \text{MVN}(\mathbf{0}, \Sigma_{\mathbf{Y}}), \\
\end{split}
\end{equation}
where $\circ$ denotes element-wise matrix multiplication, $\Sigma_{\mathbf{Y}} \in \mathbb{R}^{kq \times kq}$ is the covariance matrix induced via the KCF modeling, and $\mathbf{E}_i$ is the residual matrix. The prior distributions for components of $\mathbf{A}$, $\mathbf{Z}$, $\mathbf{M}_{i}$, and $\mathbf{E}_{i}$ have been described in Section~\ref{uncover_underlying_structure}. 

\subsection{Choosing the number of factors}
\label{choose_k}
In practice, the choice of $k$ is primarily based on previous knowledge about the data at hand. Where this is unavailable, we recommend trying several different values for $k$ and comparing results, including quantitative metrics (such as prediction performance adopted in this work) and practical interpretation. Note that it may be possible to identify when $k$ is unnecessarily large through factor loading estimates (as we will show in the simulation), since redundant factors will not have non-zero loadings. 

\section{Inference}
\label{mcem}

In this section, we describe inference for our proposed model. We develop an MCEM framework to obtain the MLE for DGP hyperparameters $\boldsymbol{\Theta}$ in Section \ref{mcem_for_gp_estimation} and the framework is summarized in Algorithm \ref{mcem_summary}. For fixed DGP hyperparameter values, we propose a Gibbs sampler for the other variables in the model, denoted by $\boldsymbol{\Omega}=\{\mathbf{M},\mathbf{Y}, \mathbf{A}, \mathbf{Z}, \boldsymbol{\rho}, \boldsymbol{\pi}, \boldsymbol{\sigma}, \boldsymbol{\phi}\}$, where $\mathbf{M}=\{\mathbf{M}_i\}_{i=1,...,n}$, $\mathbf{Y}=\{\mathbf{Y}_i\}_{i=1,...,n}$, $\boldsymbol{\rho}=\{\rho_a^2\}_{a=1,...,k}$, $\boldsymbol{\pi}=\{\pi_a\}_{a=1,...,k}$,   $\boldsymbol{\sigma}=\{\sigma_g^2\}_{g=1,...,p}$, $\boldsymbol{\phi}=\{\phi_g^2\}_{g=1,...,p}$ (Section \ref{gibbs_sample}). This sampler serves two purposes. First, within the MCEM algorithm (called ``Gibbs-within-MCEM'' hereafter), it generates samples for approximating the expectation in Equation \ref{approximated_expectation}, which is used for updating estimates of $\boldsymbol{\Theta}$. Second, after the final DGP estimate, denoted by $\widehat{\boldsymbol{\Theta}}^{\text{MLE}}$, is obtained from MCEM, we proceed with implementing the sampler (called ``Gibbs-after-MCEM'' hereafter) to find the final posterior distribution of interest, which is $f(\boldsymbol{\Omega}|\mathbf{X}, \widehat{\boldsymbol{\Theta}}_{\text{MLE}})$, where $\mathbf{X}=\{\mathbf{X}_i\}_{i=1,...,n}$ represents observed gene expression. We discuss how we handle identifiability challenges in Section \ref{identifiability}.

\subsection{MCEM 
for estimating cross-correlation determined by DGP hyperparameters}
\label{mcem_for_gp_estimation}

\subsubsection{Options for Estimating DGP Hyperparameters}
\label{options}
To estimate $\boldsymbol{\Theta}$, two strategies have been widely adopted: Fully Bayesian (FB) and Empirical Bayes (EB). The former proceeds by assigning prior distributions to $\boldsymbol{\Theta}$ to account for our uncertainty. The latter proceeds by fixing $\boldsymbol{\Theta}$ to reasonable values based on the data; for example, we can set them to the MLE. Compared to FB, EB sacrifices the quantification of uncertainty for a lower computational cost. Here, we adopt the EB approach because the key quantities of interest for inference in our model are the factor loading matrix $\mathbf{L}$, latent factors $\mathbf{Y}_i$, and the prediction of gene expression, rather than DGP hyperparameters $\boldsymbol{\Theta}$. Therefore, we simply want a reasonably good estimate of $\boldsymbol{\Theta}$ to proceed, without expending excessive computation time.

\subsubsection{Finding the MLE for DGP Hyperparameters via an MCEM Framework}
\label{mle_for_gp}
To derive $\widehat{\boldsymbol{\Theta}}^{\text{MLE}}$, we first note, with $f$ denoting probability density function,  the factorisation
        \begin{align}
        \label{eqn:marglike}
            \begin{split} 
            f(\mathbf{X},\boldsymbol{\Omega}|\boldsymbol{\Theta}) = f(\mathbf{X}|\mathbf{M}, \mathbf{Y}, \mathbf{A}, \mathbf{Z}, \boldsymbol{\phi}) f(\mathbf{M}|\boldsymbol{\sigma}) f(\mathbf{Y}|\boldsymbol{\Theta})f(\mathbf{A}|\boldsymbol{\rho})f(\mathbf{Z}|\boldsymbol{\pi})f(\boldsymbol{\phi})f(\boldsymbol{\sigma})f(\boldsymbol{\rho})f(\boldsymbol{\pi})
            \end{split}
        \end{align}
which highlights that the marginal likelihood function $f(\mathbf{X}|\boldsymbol{\Theta}) =\int{\!f(\mathbf{X},\boldsymbol{\Omega}|\boldsymbol{\Theta})}d\boldsymbol{\Omega}$ with respect to $\boldsymbol{\Theta}$ for our proposed model in Section \ref{proposed_model}
requires high-dimensional integration. 

To deal with the integration, one strategy is the Expectation-Maximization (EM) algorithm \citep{dempster1977maximum}, in which we view $\boldsymbol{\Omega}$ as hidden variables and then iteratively construct a series of estimates $\widehat{\boldsymbol{\Theta}}^{(l)}$, $l=1,2,3,...$, that converges to $\widehat{\boldsymbol{\Theta}}^{\text{MLE}}$ \citep{wu1983convergence} by alternating between E-steps and M-steps. At the $l$th iteration, the E-step requires evaluation of the ``Q-function'', which is the conditional expectation of the log-likelihood of the complete data $\{\mathbf{X}, \boldsymbol{\Omega}\}$ given the observed data $\mathbf{X}$ and the previously iterated value $\widehat{\boldsymbol{\Theta}}^{(l-1)}$.
\begin{equation}
\label{q_function}
\begin{split}
Q(\boldsymbol{\Theta},\widehat{\boldsymbol{\Theta}}^{(l-1)})&= \mathbb{E}_{\boldsymbol{\Omega}}\left[\text{ln}f(\mathbf{X}, \boldsymbol{\Omega}|\boldsymbol{\Theta})\,\middle|\,\mathbf{X}, \widehat{\boldsymbol{\Theta}}^{(l-1)}\right].
\end{split}
\end{equation}
In the M-step, this expectation is maximized to obtain the updated parameter $\widehat{\boldsymbol{\Theta}}^{(l)}
= \underset{\boldsymbol{\Theta}}{\mathrm{arg\ max}}\, Q(\boldsymbol{\Theta},\widehat{\boldsymbol{\Theta}}^{(l-1)}).
$
EM keeps updating parameters in this way until the pre-specified stopping condition is met.


As for many complex models, the analytic form of the 
conditional expectation 
in Equation \ref{q_function} is unavailable. To address this issue, a Monte Carlo version of EM (MCEM) has been developed \citep{wei1990monte,levine2001implementations
}, which uses Monte Carlo samples to approximate the exact expectation. \cite{casella2001empirical} showed that MCEM
produces consistent estimates of posterior distributions and asymptotically valid confidence sets. 

Suppose that at the $l$th iteration ($l \geq 1$), we have $R^{(l)}$ samples $\{\boldsymbol{\Omega}^{r}\}_{r = 1, \dots, R^{(l)}}$ drawn from the posterior distribution $f(\boldsymbol{\Omega}|\mathbf{X},\widehat{\boldsymbol{\Theta}}^{(l-1)})$ using an MCMC sampler (see Section~\ref{gibbs_sample}), then Equation \ref{q_function} can be approximated as,
\begin{align}
\label{approximated_expectation}
\begin{split}        Q(\boldsymbol{\Theta},\widehat{\boldsymbol{\Theta}}^{(l-1)}) \approx \widetilde{Q}(\boldsymbol{\Theta},\widehat{\boldsymbol{\Theta}}^{(l-1)})&=\frac{1}{R^{(l)}}\sum_{r=1}^{R^{(l)}}\text{ln}f(\mathbf{X}, \boldsymbol{\Omega}^{r}|\boldsymbol{\Theta}).\\
\end{split}
\end{align}
The M-Step maximizes Equation~\ref{approximated_expectation} with respect to $\boldsymbol{\Theta}$, but after applying the factorisation in Equation \ref{eqn:marglike}, the only term depending on $\boldsymbol{\Theta}$ is $f(\mathbf{Y}^{r}|\boldsymbol{\Theta})$. This means the Q-function can be simplified to
\begin{align}
\begin{split}
\widetilde{Q}(\boldsymbol{\Theta},\widehat{\boldsymbol{\Theta}}^{(l-1)})&=\frac{1}{R^{(l)}}\sum_{r=1}^{R^{(l)}}\text{ln}f(\mathbf{Y}^{r}|\boldsymbol{\Theta}).
\end{split}
\label{simplied_approximation}
\end{align}
As a result, the task has now been reduced to finding $\boldsymbol{\Theta}$ that can maximize the likelihood function of $\{\mathbf{Y}^{r}\}_{r=1,...,R^{(l)}} = \{\mathbf{Y}_{i}^{r}\}_{i=1,...,n; r=1,...,R^{(l)}}$, given that each $\mathbf{Y}_{i}^{r}$ follows a DGP distribution determined by $\boldsymbol{\Theta}$. Although gene expressions are measured at irregular time points for different individuals, samples of latent factors are available at common times $\mathbf{t}$; therefore enabling the use of GPFDA to estimate the MLE. 

The MCEM framework described above is summarized in Algorithm \ref{mcem_summary}. To provide good initial values for MCEM, we implemented a two-step approach using available software (see Supplementary B for details). Implementation challenges of MCEM are rooted in one common consideration: the computational cost of the algorithm. We discuss these challenges, including the choice of MCMC sample size $R^{(l)}$ and the stopping condition, in Supplementary A.4 and Supplementary A.5, respectively.

\subsection{Gibbs sampler for other variables under fixed GP estimates}
\label{gibbs_sample}
To acquire samples for $\boldsymbol{\Omega}$ from the posterior distribution under fixed DGP estimates $f(\boldsymbol{\Omega}|\mathbf{X},\widehat{\boldsymbol{\Theta}})$, we use a Gibbs sampler since full conditionals for variables in $\boldsymbol{\Omega}$ are analytically available. We summarise the high-level approach here; details are available in Supplementary A.1.

For the key variables $\mathbf{Z}$, $\mathbf{A}$, and $\mathbf{Y}$, we use blocked Gibbs to improve mixing. To block-update the $g$th row of the binary matrix $\mathbf{Z}$, we compute the posterior probability under $2^k$ possible values of $(Z_{g1},...,Z_{gk})$, then sample with corresponding probabilities. To update the $g$th row of the regression coefficient matrix $\mathbf{A}$,  $(A_{g1},...,A_{gk})$, we draw from a MVN distribution. When updating $\vect({\mathbf{Y}_{i}}^{T})$, the vectorized form of the $i$th individual's factor scores, we factorise $f(\vect({\mathbf{Y}_{i}}^{T})|\mathbf{X},\widehat{\boldsymbol{\Theta}},\boldsymbol{\Omega} \setminus \vect({\mathbf{Y}_{i}}^{T}))$, where $\boldsymbol{\Omega} \setminus \vect({\mathbf{Y}_{i}}^{T})$ denotes the remaining parameters excluding $\vect({\mathbf{Y}_{i}}^{T})$, according to the partition  $\vect({\mathbf{Y}_{i}}^{T}) = (\vect({\mathbf{Y}_{i,\text{obs}}^{T}}), \vect({\mathbf{Y}_{i,\text{miss}}^{T}}))$.
The first factor $f(\vect({\mathbf{Y}_{i,\text{obs}}^{T}})|\mathbf{X}_{i},\widehat{\boldsymbol{\Theta}}, \boldsymbol{\Omega} \setminus \vect({\mathbf{Y}_{i}}^{T}))$ follows a MVN distribution depending on measured gene expression.
The second factor
$f(\vect({\mathbf{Y}_{i,\text{miss}}^{T}})|\vect({\mathbf{Y}_{i,\text{obs}}^{T}}), \widehat{\boldsymbol{\Theta}})$ also follows a MVN distribution according to standard properties of the DGP model.
Our sampler samples from these two factors in turn.


\subsection{Identifiability challenges}
\label{identifiability}
A challenge of factor analysis is  identifiability, and we address non-identifiability of two types (see Supplementary A.3 for details). First, for the covariance matrix for latent factors $\Sigma_{\mathbf{Y}}$ we constrain its main diagonal elements to be $1$ (in other words, the variance of each factor at each time point is $1$). Second, to address sign-permutation identifability of the factor loadings $ \mathbf{L}$ and factor scores $\vect({\mathbf{Y}_{i}}^{T})$, we post-process samples using the R package ``factor.switch'' \citep{papastamoulis2022identifiability}.

\section{Simulation}
\label{simulation_study}

To assess the performance of our approach, we simulated gene expression observations from our model as described in Section \ref{simulation_setting}, and fitted the proposed and comparator models as described in Section \ref{analysis_methods}. Section \ref{performance_metrics} introduces the assessment metrics and Section \ref{results} describes the results.

\subsection{Simulation setting}
\label{simulation_setting}
To mimic the real data, we chose the sample size $n = 17$. The number of training and test time points was set to $q = 8$ and $u = 2$, respectively, for all individuals (i.e. we split the observed data into training and test datasets to assess models’ performance in predicting gene expression). We set the true number of latent factors as $k = 4$. 

Since recovering latent factor trajectories is one of our major interests, we considered 4 different mechanisms for generating true latent factors according to a $2 \times 2$ factorial design: the first variable determined whether different factors were actually correlated (``C'') or uncorrelated (``U''), and the second variable determined whether the variability of factors was small (``S'') or large (``L''). The mean value for each factor score $y_{ija}$ was fixed to be $0$, and we generated the covariance matrix $\Sigma_{\mathbf{Y}}$ for different scenarios in the following way: (1) under scenario ``C'', we set the cross-correlations based on the estimated covariance matrix from real data under $k=4$; otherwise under scenario ``U'' the true cross-correlations were set to $0$. (2) under scenario ``S'', we set the standard deviation for factors 1-4 to be $0.21$, $0.23$, $0.21$, and $0.17$, respectively, following the estimated results from real data under $k=4$; while under scenario ``L'', the standard deviation for each $y_{ija}$ was set to be $1$. Below we use the factors' generation mechanism to name each scenario. For example, ``scenario CS'' refers to the case where true factors were correlated (C) and had small (S) variability. 

Each factor was assumed to regulate $10\%$ of all genes, and the total number of genes was $p = 100$.  Note that we allow for the possibility that one gene may be regulated by more than one factor. If a gene was regulated by an underlying factor, then the corresponding factor loading was generated from a normal distribution $\text{N}(4,1^2)$; otherwise the factor loading was set to $0$. Each subject-gene mean $\mu_{ig}$ was generated from $\text{N}(\mu_g, \sigma^2_g)$, where $\mu_g$ ranged between $4$ and $16$ (to match the real data) and $\sigma_g = 0.5$. Finally, observed genes were generated according to Equation \ref{factor_analysis_scaler_representation}, $x_{ijg} \sim \text{N}(\mu_{ig} + \sum_{a=1}^{k} l_{ga}y_{ija}, \phi_g^2)$, where $\phi_g = 0.5$.

Note that in addition to the aforementioned setting, we also considered more simulation scenarios, to investigate the performance of our approach under different $n$, $p$, $q$ and proportion of overlap between factors. Supplementary C.2.2-C.2.5 details results under those scenarios, and below we present results only under the setting described above.

\subsection{Analysis methods}
\label{analysis_methods}
For all scenarios, we fit the data using the proposed approach BSFA-DGP under $k = 4$. In addition, we investigate how the mis-specification of $k$ impacts model performance using the scenario CS as an example (Supplementary C.2.1). Hyperparameters and other tuning parameters are detailed in Supplementary B.2.


We compared our approach to a traditional approach that models each latent factor independently, specifically assuming an IGP prior for each factor trajectory $\mathbf{y}_{ia}$ (other parts of the model remain unchanged). This model specification is referred to as BSFA-IGP. We assessed convergence of the Gibbs samples of the continuous variables from three parallel chains using Rhat \citep{gelman1992inference}. Convergence of the latent factor scores $y_{ija}$ and predictions of gene expression $x_{ijg}^{\text{new}}$ were of particular interest. We used $1.2$ as an empirical cutoff. For factor loadings $l_{ga}$, which are the product of  binary variable $Z_{ga}$ and continuous variable $A_{ga}$, we first summarized the corresponding $Z_{ga}$: if the proportion of $Z_{ga}=0$ exceeded $0.5$ for all chains,  then $l_{ga}$ was directly summarized as $0$; otherwise $l_{ga}=A_{ga}$, and we assessed convergence for these factor loadings using Rhat.

\subsection{Performance metrics}
\label{performance_metrics}

We assessed performance according to four aspects: estimation of the correlation structure; prediction of gene expression; recovery of factor trajectories; and estimating factor loadings.


To assess estimation of correlation structure, we compared the final correlation estimate with the truth and additionally, we monitored the change in the correlation estimate throughout the whole iterative process, to assess the speed of convergence of the algorithm.

We used three metrics to assess the performance in predicting gene expressions: mean absolute error ($\text{MAE}_{\mathbf{X}}$), mean width of the $95\%$ predictive interval ($\text{MWI}_{\mathbf{X}}$), and proportion of genes within the $95\%$ predictive interval ($\text{PWI}_{\mathbf{X}}$) (see Supplementary A.6 for details).

To assess the performance in recovering factor trajectories underlying gene expression observed in the training data, we first calculate the mean absolute error ($\text{MAE}_{\mathbf{Y}}$) for each scenario. In addition, we plot true and estimated factor trajectories for visual comparison. For the convenience of discussing results, we present trajectories for factor 1 of person 1 as an example. This chosen person-factor estimation result is representative of all factors of all people. 



To evaluate the ability to estimate factor loadings, we consider two perspectives. First, for a specific factor, could the model identify all genes that were truly affected by this factor? Second, for a specific gene, could the model identify all factors that regulate this gene? To answer these questions, we calculated $95\%$ credible intervals for $l_{ga}$s and presented those $l_{ga}$s of which the interval did not contain $0$ in the heatmap, using posterior median estimates. 


\subsection{Results}
\label{results}
\begin{figure}[htp]
\centering
\includegraphics[width=\textwidth]{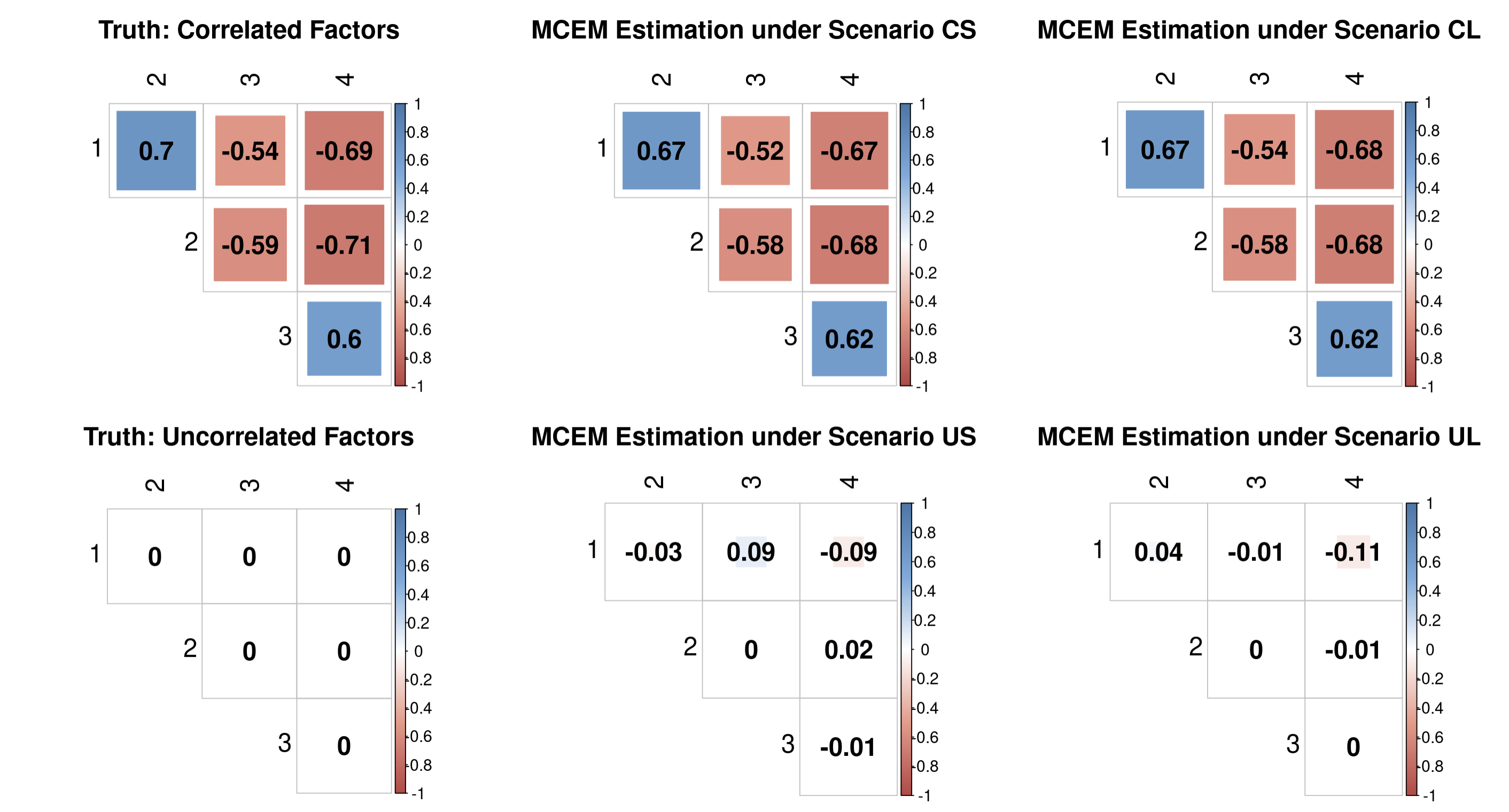}
\caption{Comparison between true and estimated cross-correlation matrices using the DGP model: all scenarios, number of latent factors $k$ correctly specified as $4$. The first column displays true correlation, while the second and third columns display estimates. The first and second row show values under correlated and uncorrelated factors, respectively. This figure appears in color in the electronic version of this article.}
\label{sim_truth_comparison_for_all_scenarios}
\end{figure}

When the number of latent factors $k$ is correctly specified, our model returns satisfactory cross-correlation estimates under all data generation mechanisms considered (Figure \ref{sim_truth_comparison_for_all_scenarios}). 
Furthermore, despite the initial (from the two-step approach) being poor, MCEM is able to propose estimates of the cross-correlations that rapidly approximate the truth (Supplementary Figure 2 shows an example under the scenario CS).  

For the inference of $x_{ijg}^{\text{new}}$, $y_{ija}$,  $l_{ga}$ using Gibbs-after-MCEM, Supplementary Table 1 lists the largest Rhat for each variable type, and reflect no apparent concerns over non-convergence. In terms of predicting gene expression, similar $\text{PWI}_{\mathbf{X}}$ is observed for both models. However, the DGP specification always leads to smaller $\text{MAE}_{\mathbf{X}}$ and narrower $\text{MWI}_{\mathbf{X}}$, even when the factors are uncorrelated in truth. This indicates more accuracy and less uncertainty in prediction (Table \ref{sim_prediction_performance_comparison_k_4}). 

Summary results in Supplementary Table 2 suggest that estimating factors is generally easier when the true variability of factors is larger. A closer inspection of factor trajectories delineated in Supplementary Figure 3 further confirms this: under scenarios CL and UL, both DGP and IGP models are able to recover the shape of factor trajectories very well. This could be explained by the relatively strong expression of latent factors. In contrast, in the case of CS and US, where expression is relatively weak, recovery of details of factor shapes is harder: if factors are not correlated at all (scenario US), both DGP and IGP fail to recover the details at the $2$nd and $5$th time point, though the overall shape is still close to the truth. However, if factors are truly correlated (scenario CS), DGP is able to recover trajectories very well, due to its ability to borrow information from other related factors. 

\begin{figure}[htp]
\centering
\includegraphics[width=\textwidth]
{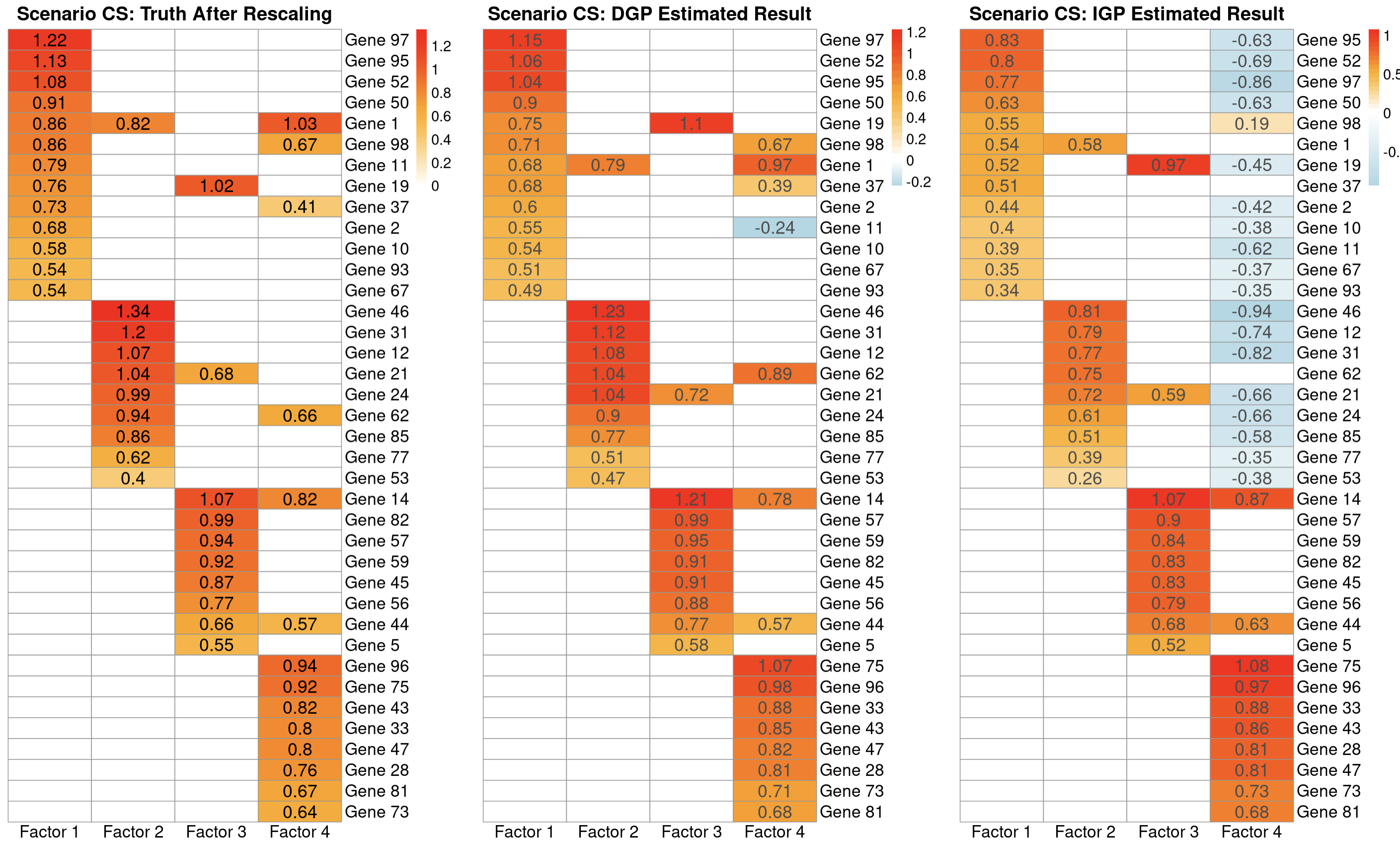}
\caption{Comparison between true and estimated factor loadings: scenario CS, with the number of latent factors $k$ correctly specified as $4$. Genes displayed in the heatmap of truth (first column) are ordered following two rules: first, genes on factors with smaller indexes are ranked first; second, genes with larger absolute factor loadings are ranked first. Genes displayed in the heatmaps of estimates (second and third columns) follow the ordering of the ground truth to facilitate comparison. This figure appears in color in the electronic version of this article.}
\label{estimated_factor_loading_matrices_for_scenario_fcs_rescale}
\end{figure}

With regard to factor loading estimation, Figure \ref{estimated_factor_loading_matrices_for_scenario_fcs_rescale} shows results under the scenario CS, and Supplementary Figures 4,5 and 6 show results for the remaining scenarios. Both DGP and IGP perform well at identifying the correct genes for a given factor under all scenarios. Both models also perform well  at identifying the correct factors for a given gene under scenarios CL, UL and US, but under the scenario CS we observe that DGP still performs well whereas IGP performs less well. As can be seen from Figure \ref{estimated_factor_loading_matrices_for_scenario_fcs_rescale}, IGP specification leads to the result that genes estimated to be significantly loaded on the first and second factor also significantly load on the fourth factor. One possible explanation for this is that, if in truth, factors have strong signals and/or are not correlated at all (as in scenarios CL, UL and US), then it is relatively easy to distinguish contributions from different factors. In contrast, under the scenario CS where factors actually have weak signals and are highly-correlated (true correlation between factor 1 and 4 is $-0.69$, and correlation between factor 2 and 4 is $-0.71$, as can be found from Figure \ref{sim_truth_comparison_for_all_scenarios}), it is more difficult. DGP specification explicitly takes the correlation among factors into consideration, allowing improved estimation.

\section{Data application}
\label{data_application}

We ran $500,000$ iterations for the final Gibbs sampler, with a $50\%$ burn-in proportion and retained only every $100$th iteration. We also compared our results with two alternative models, both of which assume independence between factors: the BSFA-IGP model, and the previous model in \cite{chen2011predicting}, which adopted spline functions to model each factor trajectory independently.

\subsection{Statistical results}
We fitted the model with $k=2, 3, \dots, 10$ factors. We display results under $k = 5$ after a comprehensive consideration of biological interpretation, computational cost and statistical prediction performance. Biologically, we found that $2$ biological pathways (immune response pathway and ribosome pathway; see Section~\ref{biological_interpretation} for details) were consistently identified when $k$ ranged from $3$ to $10$ while $k = 2$ led to only $1$ pathway (immune response pathway) identified. Computationally, once $k$ was larger than $5$, our algorithm became slower as the parameter space was very large. Statistically, among the remaining models, $k = 5$ showed the best prediction performance (see Supplementary Table 6).

In terms of the prediction performance, similar $\text{PWI}_{\mathbf{X}}$ is observed under BSFA-DGP ($0.951$) and BSFA-IGP ($0.956$); whereas $\text{MAE}_{\mathbf{X}}$ and $\text{MWI}_{\mathbf{X}}$ are both smaller under BSFA-DGP ($\text{MAE}_{\mathbf{X}} \ 0.211$; $\text{MWI}_{\mathbf{X}} \ 1.113$) than BSFA-IGP ($\text{MAE}_{\mathbf{X}} \ 0.217$; $\text{MWI}_{\mathbf{X}} \ 1.213$). This again demonstrates the advantage of reducing prediction error and uncertainty if cross-correlations among factors are taken into consideration. 

\begin{figure}[htp]
\centering
\includegraphics[width=\textwidth]
{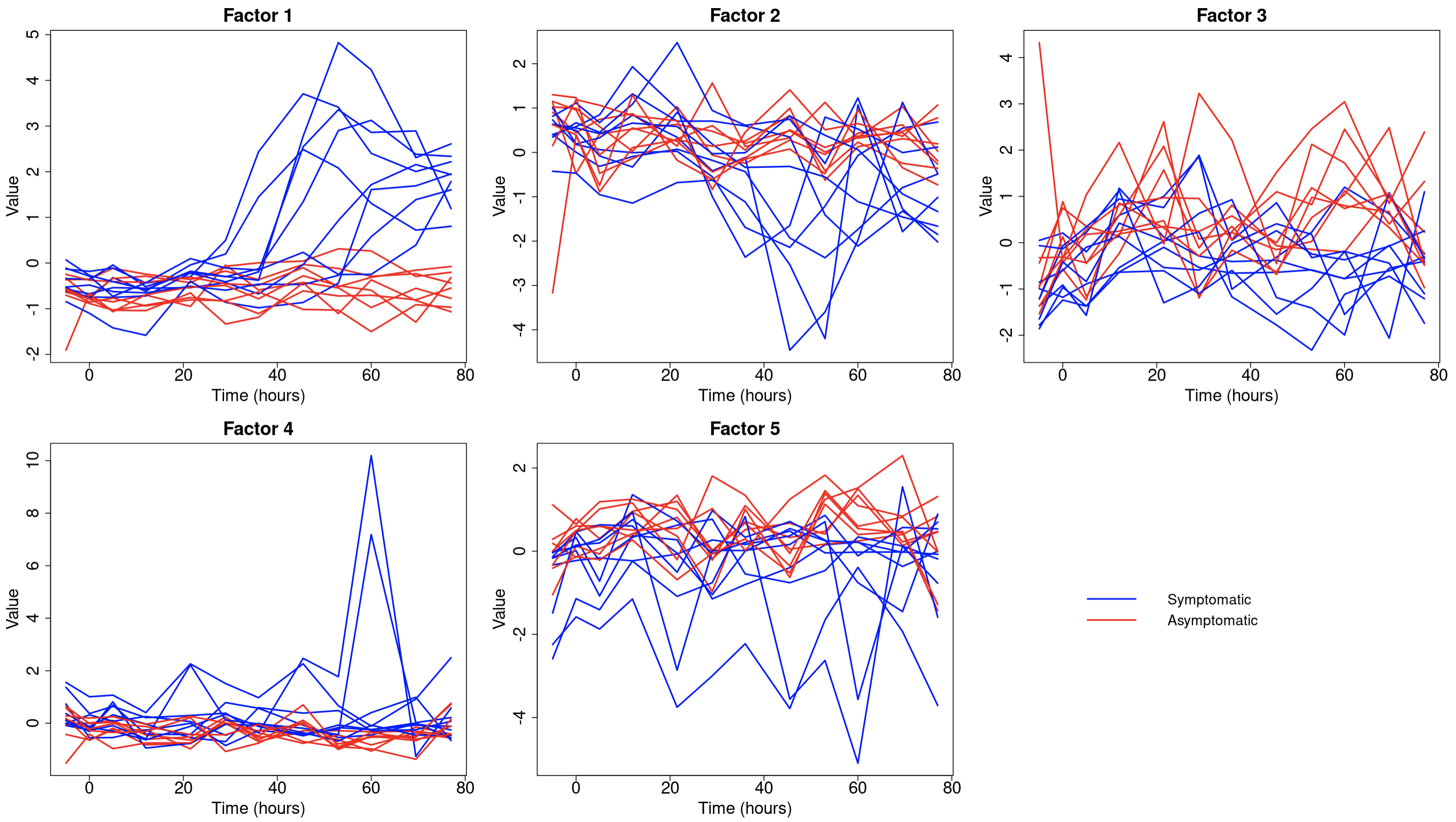}
\caption{Estimated trajectories for all factors in the H3N2 data. Each panel displays all subjects' trajectories for a single factor. This figure appears in color in the electronic version of this article.}
\label{bsfadgp_all_factors}
\end{figure}

Estimated factor trajectories using BSFA-DGP are displayed in Figure \ref{bsfadgp_all_factors}. Among all, factor $1$ is able to distinguish symptomatic people from asymptomatic people most clearly, and we find that its shape is largely similar to the ``principal factor'' identified in \cite{chen2011predicting}. Despite the similarity, it is noteworthy that the trajectory of our factor $1$ is actually more individualized and informative than their principal factor (see Supplementary D.2 for discussion in detail). Compared with BSFA-IGP, our model BSFA-DGP reduces the variability within subject-specific trajectory, and consequently it slightly increases the difference between the symptomatic and asymptomatic. Although results look similar at first glance (Supplementary Figure 17), a more careful inspection reveals that BSFA-DGP has a shrinkage effect compared to BSFA-IGP. Take factor 1 as an example (similar observations for other factors). For the blue line at the bottom: at $38$ hours, it remains around $-1$ under BSFA-DGP while reaches as low as $-2$ under BSFA-IGP, making it harder to differentiate this symptomatic subject from asymptomatic subjects with red lines. 

\subsection{Biological intepretation}
\label{biological_interpretation}
To identify the biological counterpart (i.e., pathway) of the statistical factor, we used the KEGG Pathway analysis of the online bioinformatics platform \href{https://david.ncifcrf.gov/home.jsp}{DAVID}. We uploaded the total $11,961$ genes as the ``Background'' and the top $50$ genes loaded on each factor as the ``Gene List''. 
This showed that factor 1 can be interpreted as `immune response pathway' and factor 3 as `ribosome pathway', while other factors have no known biological meaning (see Supplementary D.4). In addition, our MCEM algorithm estimates the magnitude of the cross-correlation coefficient to be $0.57$ between factors 1 and 3. This statistically moderate but biologically noteworthy correlation between these two pathways is consistent with existing biological knowledge \citep[e.g. see][which concludes that ribosome biogenesis restricts innate immune responses to virus infection]{bianco2019ribosome}. This interrelated relationship would be missed by an approach that assumed independence between pathways, such as BSFA-IGP.

\section{Discussion}
\label{discussion}

In this paper we propose a BSFA-DGP model, which is the first attempt to relax the classical assumption of independent factors for longitudinal data. By borrowing information from correlated factor trajectories, this model has demonstrated advantages in estimating factor loadings, recovering factor trajectories and predicting gene expression when comparing with other models. 

We also develop an MCEM algorithm for the inference of the model; the main motivation for adopting this algorithm is that it can make full use of the existing package to estimate DGP hyperparameters. In the M-step of MCEM, GPFDA can be directly implemented to find the maximizer of the Q-function; thus we can exploit the existing optimization algorithm. In addition to its implementation convenience, the modular structure of the MCEM framework makes it generalizable to other complex models involving the DGP model component. For example, when using a DGP model for the latent continuous variable introduced to model multivariate dependent, non-continuous data (such as binary or count data), \cite{shi2017regression} used Laplace Approximation; the MCEM framework we develop would be an alternative inference approach in this case.

Computation time of our approach is largely dependent on GPFDA, which returns hyperparameter estimates for the DGP to maximize the exact likelihood of observing the inputs $\mathbf{Y}$. The computational complexity of GPFDA at the $l$th iteration of MCEM is $O(nR^{(l)}k^2q^2)$, indicating the computational cost scales linearly with the sample size $n$ and the current number of MCMC samples $R^{(l)}$, quadratically with the number of latent factors $k$ and the number of observed time points $q$. MCEM scales well with $p$, as it is not involved in the expression of complexity at all; it took around $1.5$ hours for the real data in Section \ref{data_application} on a standard laptop (Quad-Core Intel Core i5). Potential approaches that can further improve the computational performance include a model approximation method proposed in \cite{alvarez2010efficient}, or stochastic variational inference, as discussed in \cite{hensman2013gaussian}.


Throughout this paper, the number of latent factors $k$ was fixed and needed to be pre-specified. Though results of our model could infer redundant factors (as demonstrated in the simulation study), an automatic approach to infer this number from the data may be preferable in some contexts. A potential solution would be to introduce the Indian Buffet Process as a prior distribution over equivalence classes of infinite-dimensional binary matrices \citep{knowles2011nonparametric}. In addition, although our empirical evidence suggests that the constraints we adopt are sufficient to avoid non-identifiability, theoretical investigation would be helpful in future work. Extensions of our methodology to handle non-normal data, limits of detection, covariates and batch effects would also be interesting avenues for future research. Furthermore, in some settings it may be helpful to explicitly incorporate the clinical outcome into the model, which could be another direction for future research. 

\section{Software}

The R code used for implementing the proposed BSFA-DGP model is available as an R package, DGP4LCF, on Github: \url{https://github.com/jcai-1122/DGP4LCF}. The package contains vignettes, which illustrate the usage of the functions within the package by applying them to analyze simulated dataset. The release used in this paper is available at https://doi.org/10.5281/zenodo.8108150. 

\section*{Acknowledgments}
This work is supported through the United Kingdom Medical Research Council programme grants \texttt{MC\_UU\_00002/2}, \texttt{MC\_UU\_00002/20}, \texttt{MC\_UU\_00040/02} and \texttt{MC\_UU\_00040/04}. The authors are grateful to Oscar Rueda for the helpful discussion on the bioinformatics side of the project. 

{\it Conflict of Interest}: None declared.


\bibliography{refs}

\clearpage 

\SetKwComment{Comment}{/* }{ */}

\begin{algorithm}[hbt!]
\caption{MCEM Algorithm for the MLE of DGP Hyperparameters.} \label{mcem_summary}
\SetKwInOut{KwIn}{Input}
\SetKwInOut{KwOut}{Output}

\KwIn{Observed gene expression $\mathbf{X}_i$ and time points $\mathbf{t}_i$, $i = 1,...,n$ for $n$ people.}
\KwOut{The MLE of DGP hyperparameters $\widehat{\boldsymbol{\Theta}}^{\text{MLE}}$.}

\textbf{Step 1: Initialization Step}\\

 \begin{itemize}
 \item Center $\mathbf{X}_i$ to obtain $\mathbf{X}_i^{c}$ (as defined in Section \ref{analysis_methods}), which is input to BFRM for $\mathbf{Y}^{0}$.
 \item Input $\mathbf{Y}^{0}$ to GPFDA for $\widehat{\boldsymbol{\Theta}}^{(0)}$, and construct $\widehat{\Sigma}_\mathbf{Y}^{(0)}$ using $\widehat{\boldsymbol{\Theta}}^{(0)}$ and $\mathbf{t} = \bigcup\limits_{i=1}^{n}\mathbf{t}_{i}$ .
 \item Initialize the starting sample size $R^{(1)}$; the counter $w=0$, which counts the total number of sample size increase, and specify its upper limit as $W$, as described in Supplementary A.5.
 \end{itemize}

\textbf{Step 2: Iteration Step, Starting from $l=1$}

\SetKwRepeat{Do}{do}{while}

\Do{$w \leq W$}{

\begin{enumerate}[2.1]
\item Draw $R^{(l)}$ samples of $\boldsymbol{\Omega}$ from the Gibbs sampler $f(\boldsymbol{\Omega}|\mathbf{X}, \widehat{\Sigma}_\mathbf{Y}^{(l-1)})$ in Section \ref{gibbs_sample}, where $\widehat{\Sigma}_\mathbf{Y}^{(l-1)}$ is constructed by $\widehat{\boldsymbol{\Theta}}^{(l-1)}$ and $\mathbf{t}$ . 
\item Retain $R^{(l)}_{\text{remain}}$ samples by discarding burn-in and thinning.
\item Align post-processed samples $\{\mathbf{L}^{r}, \mathbf{Y}_i^{r}\}_{r=1,...,R^{(l)}_{\text{remain}}}$ using R package ``factor.switch''.
\item Input aligned samples $\{\mathbf{Y}_i^{r}\}_{r=1,...,R^{(l)}_{\text{remain}}}$ to GPFDA to obtain $\widehat{\Sigma}_\mathbf{Y}^{(l)}$.
\item Calculate the lower bound proposed by \cite{caffo2005ascent} using $\widehat{\Sigma}_\mathbf{Y}^{(l-1)}$, $\widehat{\Sigma}_\mathbf{Y}^{(l)}$ and $\{\mathbf{Y}_i^{r}\}_{r=1,...,R^{(l)}_{\text{remain}}}$ (see Supplementary A.4.2 for details). 
\end{enumerate}
 \eIf{$\text{LB} > 0$}{
   $R^{(l+1)}\gets R^{(l)}$\;
   $l\gets l+1$\;
 }{
   $R^{(l)}\gets R^{(l)}+\frac{R^{(l)}}{m}$\;
   $w \gets w+1$\;
 }
}

\end{algorithm}

\clearpage 

\begin{table}[hbt!]
  \caption{Prediction performance for all scenarios: number of latent factors $k$ correctly specified as $4$ for both DGP and IGP models. $\text{MAE}_{\mathbf{X}}$, $\text{MWI}_{\mathbf{X}}$ and $\text{PWI}_{\mathbf{X}}$ are short for mean absolute error, mean width of the $95\%$ predictive intervals, and the proportion of genes within $95\%$ the predictive intervals, respectively.}
  
 \begin{center}
    \begin{tabular}{cccccccc}
    \hline
    \multicolumn{2}{c}{Factor Generation Mechanism} & \multicolumn{3}{c}{DGP, $k=4$} & \multicolumn{3}{c}{IGP, $k=4$}\\
    \hline
     Correlation & Variability & $\text{MAE}_{\mathbf{X}}$ & $\text{MWI}_{\mathbf{X}}$ & $\text{PWI}_{\mathbf{X}}$ & $\text{MAE}_{\mathbf{X}}$ & $\text{MWI}_{\mathbf{X}}$ & $\text{PWI}_{\mathbf{X}}$\\
    \hline
     Correlated & Large & 1.29 & 6.52 & 0.95 & 1.42 & 7.85 & 0.95\\
      & Small & 0.53 & 2.66 & 0.95 & 0.55 & 2.76 & 0.95\\
     Uncorrelated & Large & 1.44 & 7.28 & 0.95& 1.53 & 7.70 & 0.95\\
     & Small & 0.55 & 2.78 & 0.95 & 0.56 & 2.79 & 0.95\\
     \hline
     
    \end{tabular}
   \end{center}
\label{sim_prediction_performance_comparison_k_4}
\end{table}

\clearpage

\renewcommand*{\algorithmcfname}{Supplementary Algorithm}

\renewcommand{\figurename}{Supplementary Figure}
\renewcommand{\tablename}{Supplementary Table}

\setcounter{figure}{0} 
\setcounter{table}{0} 

\counterwithout{equation}{section}

\clearpage

{\centering{\huge {\textbf {Supplementary Materials}} }}

\section*{A. Mathematical derivations}
\subsection*{A.1 Full conditionals of the Gibbs sampler for the proposed model}


Throughout the following derivation, ``$\circ$'' denotes element-wise multiplication, ``$-$'' denotes observed data and all parameters in the model other than the parameter under derivation, ``$||\mathbf{z}||^2$'' denotes the sum of squares of each element of the vector $\mathbf{z}$, ``$\text{diag}(\mathbf{z})$'' denotes a diagonal matrix with the vector $\mathbf{z}$ as its main diagonal elements, ``$\text{MVN}(\mathbf{z}; \boldsymbol{\mu}_{\mathbf{z}}, \Sigma_{\mathbf{z}})$'' denotes that $\mathbf{z}$ follows a multivariate normal distribution with mean $\boldsymbol{\mu}_{\mathbf{z}},$ and variance $\Sigma_{\mathbf{z}}$, and similar interpretations apply to other distributions. ``$\mathbf{z}^{T}$'' denotes the transpose of the vector or matrix $\mathbf{z}$, and ``pos'' is short for `posterior probability'.

\begin{itemize}
    \item Full conditional for the latent factors $\vect({\mathbf{Y}_{i}}^{T}),i=1,\ldots,n$
    \begin{align*}
    \begin{split}
        f(\vect({\mathbf{Y}_{i}}^{T})|-)
        & =  f(\vect({\mathbf{Y}_{i,\text{obs}}^{T}})|-) \cdot     f(\vect({\mathbf{Y}_{i,\text{miss}}^{T}})|\vect({\mathbf{Y}_{i,\text{obs}}^{T}}),\Sigma_\mathbf{Y})\\
    \end{split}
    \end{align*}
    We first sample for $\vect({\mathbf{Y}_{i,\text{obs}}^{T}})$:
    \begin{align*}
    \begin{split}
f(\vect({\mathbf{Y}_{i,\text{obs}}^{T}})|-)
        & \propto \text{MVN}(\vect({\mathbf{X}_i^{T}}); \  \vect({\mathbf{M}_i^{T}})+ \mathbf{L}_i^{*} \vect({\mathbf{Y}_{i,\text{obs}}^{T}}), \ \Sigma_{\mathbf{X}_i})
        \cdot \text{MVN}(\vect({\mathbf{Y}_{i,\text{obs}}^{T}}); \ \mathbf{0},\ \Sigma_{\mathbf{Y}_{i,\text{obs}}}) \\
        &=\text{MVN}(\mu_{\mathbf{Y}_{i,\text{obs}}}^{\text{pos}}, \ \Sigma_{\mathbf{Y}_{i, \text{obs}}}^{\text{pos}}),
    \end{split}
    \end{align*}
     with
    \begin{align*}
      \begin{split}
        \Sigma^{pos}_{\mathbf{Y}_{i,\text{obs}}}&=[{\mathbf{L}_i^{*}}^{T}\Sigma_{\mathbf{X}_{i}}^{-1}\mathbf{L}_i^{*}+\Sigma_{\mathbf{Y}_{i,\text{obs}}}^{-1}]^{-1} \\
    \mu^{\text{pos}}_{\mathbf{Y}_{i,\text{obs}}}&=\Sigma^{\text{pos}}_{\mathbf{Y}_{i,\text{obs}}} ({\mathbf{L}_i^{*}}^{T}\Sigma_{\mathbf{X}_{i}}^{-1}(\vect({\mathbf{X}_i^{T}})-\vect({\mathbf{M}_i^{T}}))).
      \end{split}
    \end{align*}
    Then we sample for $\vect({\mathbf{Y}_{i,\text{miss}}^{T}})$ from $f(\vect({\mathbf{Y}_{i,\text{miss}}^{T}})|\vect({\mathbf{Y}_{i,\text{obs}}^{T}}),\ \Sigma_\mathbf{Y})$, a MVN distribution due to the property of the DGP model \citep{shi2011gaussian}. \\
    
    In the above equations: 
    
    \begin{itemize}
    
     \item  $\Sigma_\mathbf{Y}$ is the covariance matrix of factor scores at full time $\mathbf{t}$, and $\Sigma_{\mathbf{Y}_{i,\text{obs}}}$ is a sub-matrix of it (at subject-specific time points);
     
     \item  
    $\mathbf{L}_i^{*}$ is constructed using components of the factor loading matrix $\mathbf{L}$:  $\mathbf{L}_i^{*}=(\mathbf{L}_1^{*},...,\mathbf{L}_p^{*})^{T} \in \mathbb{R}^{pq_i \times kq_i}$,  where $(\mathbf{L}_g^{*})^{T}=({\text{diag}(l_{g1})}_{q_i \times q_i},...,{\text{diag}(l_{gk})}_{q_i \times q_i}) \in \mathbb{R}^{q_i \times kq_i}$; 
    
    \item 
    
    $\Sigma_{\mathbf{X}_i}=\text{diag}(({\phi_1^2})_{\times q_i},...,({\phi_p^2})_{\times q_i}) \in \mathbb{R}^{pq_i \times pq_i}$, where $\ ({\phi_a^2})_{\times q_i}$ represents a $q_i$-dimensional row vector consisting of the scalar ${\phi_a^2}$.
    
    \end{itemize}
    
    Note that when coding the algorithm, there is no need to directly create the $pq_i \times pq_i$ diagonal matrix $\Sigma_{\mathbf{X}_{i}}^{-1}$ as the memory will be exhausted. To calculate the term ${\mathbf{L}_i^{*}}^{T}\Sigma_{\mathbf{X}_{i}}^{-1}$, we can use the property of multiplication of a diagonal matrix: post-multiplying a diagonal matrix is equivalent to multiplying each column of the first matrix by corresponding elements in the diagonal matrix. \\
    
           
           %
           
          \item Full conditional for the binary matrix $\mathbf{Z}$\\ 
          Let $\mathbf{Z}_{g \cdot} = (Z_{g1},...,Z_{gk})$ denote the $g$th row of the matrix $\mathbf{Z}, \ g=1,...,p$; then, 
          \begin{align*}
          \begin{split}
               f(\mathbf{Z}_{g \cdot}|-) 
               & \propto \prod_{i=1}^{n}\text{MVN}(\mathbf{x}_{ig}; \ \boldsymbol{\mu}_{ig} + (\mathbf{A}_{g \cdot} \circ \mathbf{Z}_{g \cdot})\mathbf{Y}_{i}, \ \text{diag}(\phi_g^{2}, \ q_i)) \cdot \prod_{a=1}^{k} \text{Bernoulli}(Z_{ga}; \pi_a),
          \end{split}
          \end{align*}
          We calculate the posterior probability under $2^k$ possible values of $\mathbf{Z}_{g \cdot}$ based on the above formula, then sample with corresponding probability.
          
          
          
          \item Full conditional for the regression coefficient matrix $\mathbf{A}$ \\
          Let $\mathbf{A}_{g \cdot} = (A_{g1},...,A_{gk})$ denote the $g$th row of the matrix $\mathbf{A}, \ g=1,...,p$; then, 
          \begin{align*}
          \begin{split}
               f(\mathbf{A}_{g \cdot}|-) 
               & \propto \prod_{i=1}^{n}\text{MVN}(\mathbf{x}_{ig}; \ \boldsymbol{\mu}_{ig} + (\mathbf{A}_{g \cdot} \circ \mathbf{Z}_{g \cdot})\mathbf{Y}_{i}, \ \text{diag}(\phi_g^{2}, \ q_i)) \cdot \text{MVN}(\mathbf{A}_{g \cdot}; \ \mathbf{0},\ \text{diag}(\boldsymbol{\rho}^2)) \\
               & =\text{MVN}(\mathbf{A}_{g \cdot}; \ \mu_{\mathbf{A}_{g \cdot}}^{\text{pos}}, \ \Sigma_{\mathbf{A}_{g \cdot}}^{\text{pos}}),
          \end{split}
          \end{align*}
         
          where 
          \begin{align*}
              \begin{split}
              \Sigma_{\mathbf{A}_{g \cdot}}^{\text{pos}}&=({\frac{\text{diag}(\mathbf{Z}_{g\cdot})(\sum_{i=1}^{n}{\mathbf{Y}_{i}}^{T}\mathbf{Y}_{i})\text{diag}(\mathbf{Z}_{g\cdot})}{\phi^2_g}+\text{diag}(\frac{1}{\boldsymbol{\rho^2}}))}^{-1} \\
              \mu_{\mathbf{A}_{g \cdot}}^{\text{pos}}&=\frac{\Sigma_{\mathbf{A}_{g \cdot}}(\text{diag}(\mathbf{Z}_{g\cdot})\sum_{i=1}^{n}{\mathbf{Y}_{i}}(\mathbf{x}_{ig}-\boldsymbol{\mu}_{ig}))}{\phi^2_g}.
              \end{split}
          \end{align*}
          and $\boldsymbol{\rho^2}=(\rho^2_1,...,\rho^2_a)$.

          \item Full conditional for the intercept $\mu_{ig}, i=1,...,n; g=1,...,p$
          \begin{align*}
          \begin{split}
              f(\mu_{ig}|-)
              &\propto \prod_{j=1}^{q_i} \text{N}(x_{ijg}; \mu_{ig} + \sum_{a=1}^{k}l_{ga}y_{ija}, \phi^2_g) \cdot \text{N}(\mu_{ig};\mu_g, \sigma^2_g) \\
              & = \text{N}(\mu_{ig}; \mu_{ig}^{\text{pos}}, \ \sigma_{ig}^{2,\text{pos}}),
          \end{split}
          \end{align*}
          where 
          \begin{align*}
          \begin{split}
             \sigma_{ig}^{2,\text{pos}} & = {(\frac{1}{\sigma^2_g} + \frac{q_i}{\phi^2_g})}^{-1} \\
             \mu_{ig}^{\text{pos}} & = (\frac{\mu_g}{\sigma^2_g} + \frac{\sum_{j=1}^{q_i}(x_{ijg}-\sum_{a=1}^{k}l_{ga}y_{ija})}{\phi^2_g}) \cdot \sigma_{ig}^{2,\text{pos}}
          \end{split}
          \end{align*}
         
          \item Full conditional for $\pi_a,\ a=1,...,k$ 
          \begin{align*}
              \begin{split}
              f(\pi_a|-)
              & \propto \prod_{a=1}^{k}\text{Bernoulli}(Z_{ga}; \pi_a) \cdot \text{Beta}(\pi_a; c_0, d_0)\\
              &=\text{Beta}(c_0+\sum_{g=1}^{p}Z_{ga},\ d_0+\sum_{g=1}^{p}(1-Z_{ga})) \\
              \end{split}
          \end{align*}
          
          \item Full conditional for $\rho_a^2,\ a=1,...,k$ 
           \begin{align*}
              \begin{split}
                f(\rho_a^2|-)
                & \propto \prod_{g=1}^{p} \text{N}(A_{ga};0, \rho^2_a) \cdot \text{Inverse-Gamma}(\rho_a^2; c_1, d_1) \\
                & =\text{Inverse-Gamma}(c_1+\frac{p}{2},\ d_1+\frac{1}{2}\sum_{g=1}^{p}A_{ga}^2) \\
              \end{split}
          \end{align*}
          
          \item Full conditional for $\sigma_g^2,\ g=1,...,p$
          \begin{align*}
          \begin{split}
            f(\sigma_g^2|-) 
            & \propto \prod_{i=1}^{n} \text{N}(\mu_{ig};\mu_g, \sigma^2_g) \cdot \text{Inverse-Gamma}(\sigma^2_g; c_2, d_2) \\
            &=\text{Inverse-Gamma}(c_2+\frac{1}{2}n, \ d_2+\frac{1}{2}\sum_{i=1}^{n}{(\mu_{ig}-\mu_g})^2)
          \end{split}
          \end{align*}

          \item Full conditional for $\phi_g^2,\ g=1,...,p$
          \begin{align*}
          \begin{split}
            f(\phi_g^2|-) 
            & \propto \prod_{i=1}^{n}\text{MVN}(\mathbf{x}_{ig}; \ \boldsymbol{\mu}_{ig} + (\mathbf{A}_{g \cdot} \circ \mathbf{Z}_{g \cdot})\mathbf{Y}_{i}, \ \text{diag}(\phi_g^{2}, \ q_i)) \cdot \text{Inverse-Gamma}(\phi^2_g; c_3, d_3) \\
            &=\text{Inverse-Gamma}(c_3+\frac{1}{2}\sum_{i=1}^{n} q_{i}, \ d_3+\frac{1}{2}\sum_{i=1}^{n}{||\mathbf{x}_{ig}-\boldsymbol{\mu}_{ig}-(\mathbf{A}_{g\cdot} \circ \mathbf{Z}_{g\cdot})\mathbf{Y}_i||}^2)
          \end{split}
          \end{align*}
           

           \item Full conditional for predictions of gene expression (only implemented when assessing models' prediction performance on the test dataset)
           
           Suppose that $\mathbf{X}_i^{\text{new}}, \mathbf{Y}_i^{\text{new}}$ represent predicted gene expression and factor expression of the $i$th individual at new time points, respectively. The posterior predictive distribution under MCEM-algorithm-returned $\widehat{\boldsymbol{\Theta}}^{\text{MLE}}$ can be expressed as 
\begin{align*}
\begin{split}
    f(\mathbf{X}_i^{\text{new}}|\ \widehat{\boldsymbol{\Theta}}^{\text{MLE}}, \boldsymbol{\Omega}) 
    & = \int f(\mathbf{X}_i^{\text{new}}, \mathbf{Y}_i^{\text{new}}|\ \widehat{\boldsymbol{\Theta}}^{\text{MLE}}, \boldsymbol{\Omega}) d \mathbf{Y}_i^{\text{new}} \\
    & = \int f(\mathbf{X}_i^{\text{new}}|\mathbf{Y}_i^{\text{new}}, \boldsymbol{\Omega}) \cdot f(\mathbf{Y}_i^{\text{new}}| \widehat{\boldsymbol{\Theta}}^{\text{MLE}},\mathbf{Y}_{i,\text{obs}}) d \mathbf{Y}_i^{\text{new}},
\end{split}
\end{align*}
where the first term of the integrand is a MVN because of the assumed factor model, and the second term is also a MVN because of the assumed DGP model on latent factor trajectories \citep{shi2011gaussian}. Therefore, once a sample of parameters $\boldsymbol{\Omega}^{r}$ is generated, the $r$th sample of ${\mathbf{Y}_i^{\text{new}}}$ can be generated from $f(\mathbf{Y}_i^{\text{new}}| \ \widehat{\boldsymbol{\Theta}}^{\text{MLE}},\mathbf{Y}_{i,\text{obs}}^{r})$, then the $r$th sample of ${\mathbf{X}_i^{\text{new}}}$ can be sampled from $f(\mathbf{X}_i^{\text{new}}|\ {\mathbf{Y}_i^{\text{new}}},\boldsymbol{\Omega}^{r})$.

\end{itemize}

\subsection*{A.2 Kernel convolution framework to model dependent Gaussian processes}
\subsubsection*{A.2.1 Illustration using two processes}

The KCF constructs correlated processes by introducing common ``base processes''. Take two processes ${y}_{a}(t)$ and ${y}_{b}(t)$ as an example. KCF constructs them as, 
\begin{equation*}
\begin{split}
\label{factor_score_modeling}
y_{a}(t)& =\eta_a(t)+\xi_a(t)+\epsilon_a(t), \\
y_{b}(t)& =\eta_b(t)+\xi_b(t)+\epsilon_b(t),
\end{split}
\end{equation*}
where $\epsilon_a(t), \ \epsilon_b(t)$ are residual errors from 
$\text{N}(0, \psi^2)$, and $\eta_a(t), \ \eta_b(t), \ \xi_a(t), \ \xi_b(t)$ are processes constructed in the following way, illustrated in Supplementary Figure~\ref{kernel_convolution} below. 

\begin{figure}[hbt!]
\centering
\includegraphics[width=0.8\textwidth]{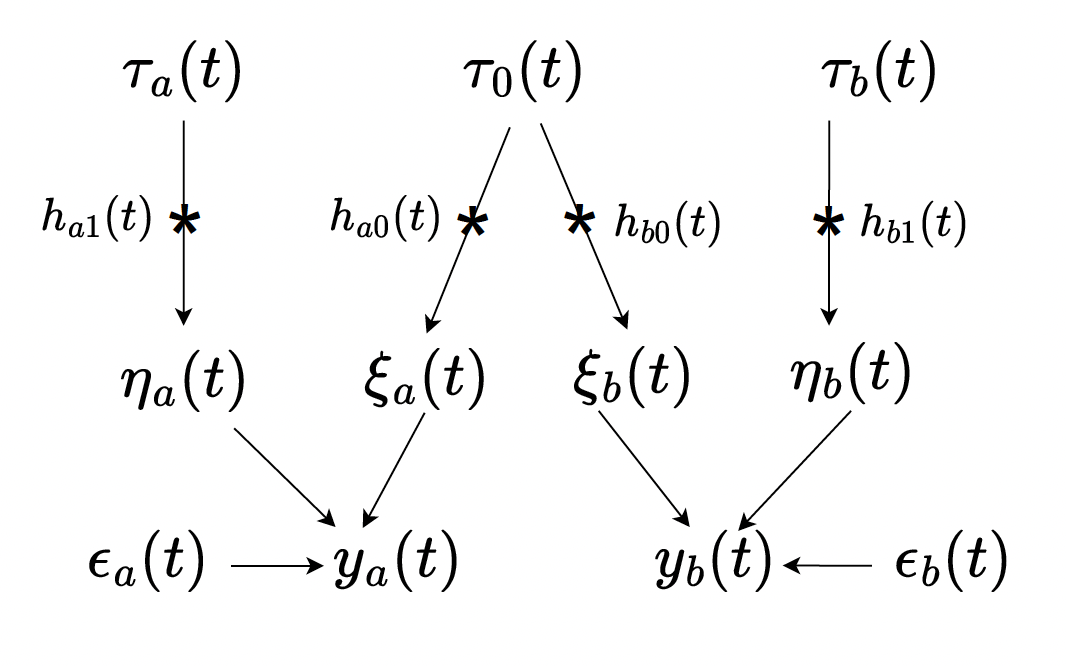}
\caption{Illustration of the kernel convolution framework for DGPs. The star ($\ast$) denotes a convolutional operation and directed arrows indicate direct dependence. $t$ denotes time; $a,b$ are indexes of factor trajectories; $\tau_a(t),\tau_0(t),\tau_b(t)$ are independent Gaussian white noise processes; $h_{a1}(t),h_{a0}(t),h_{b0}(t),h_{b1}(t)$ are Gaussian kernel functions; $\epsilon_a(t),\epsilon_b(t)$ are residuals; $y_{a}(t),y_{b}(t)$ are the $a$th and $b$th factor trajectories, respectively.}
\label{kernel_convolution}
\end{figure}

First, three independent, zero-mean base processes $\tau_0(t)$, $\tau_a(t)$ and $\tau_b(t)$ are introduced, which are all Gaussian white noise processes. The first process $\tau_0(t)$ is shared by both ${y}_{a}(t)$ and ${y}_{b}(t)$, thereby inducing dependence between them. Whereas $\tau_a(t)$ and $\tau_b(t)$ are specific to ${y}_{a}(t)$ and ${y}_{b}(t)$, respectively; they are responsible for capturing the unique aspects of each process.

Second, Gaussian kernel functions $h_{a0}(t),\ h_{a1}(t),\ h_{b0}(t),\ h_{b1}(t)$ are applied to convolve the base processes: with $h_{-0}(t)$ applied to the shared process $\tau_0(t)$ and $h_{-1}(t)$ to the output-specific processes $\tau_a(t)$ and $\tau_b(t)$,
\begin{align*}
 \xi_a(t)&=h_{a0}(t)*\tau_{0}(t), & \eta_a(t)&=h_{a1}(t)*\tau_a(t), \\
  \xi_b(t)&=h_{b0}(t)*\tau_{0}(t), & \eta_b(t)&=h_{b1}(t)*\tau_b(t),
\end{align*}
where the convolution operator $*$ is defined as $h(t) * \tau(t) = \int^{\infty}_{-\infty} h(t-s)\tau(s)ds $. All kernel functions $h(t)$ take the form $h(t) = v\,\text{exp}\{-\frac{1}{2}{B}t^2\}$, where $v$ and ${B}$ are positive parameters that are specific to each kernel function.


\subsubsection*{A.2.2 Specific form of covariance function}

Under the kernel convolution framework, the covariance function between the time $t_j$ and $t_\ell$ within a single process $a$, denoted as $C_{aa}^{Y}(t_j, t_\ell)$, can be decomposed as,
\begin{align*}
    \label{auto_covariance_expression}
    \begin{split}
         C_{aa}^{Y}(t_j,t_\ell)&=C_{aa}^{\xi}(t_j,t_\ell)+C_{aa}^{\eta}(t_j,t_\ell)+\delta_{j\ell}\psi^2,\\
        C_{aa}^{\xi}(t_j,t_\ell)&=v_{a0}^2\frac{(\pi)^{\frac{1}{2}}}{\sqrt{|B_{a0}|}}\text{exp}\{-\frac{1}{4}B_{a0}d_t^2\}, \\
        C_{aa}^{\eta}(t_j,t_\ell)&=v_{a1}^2\frac{(\pi)^{\frac{1}{2}}}{\sqrt{|B_{a1}|}}\text{exp}\{-\frac{1}{4}B_{a1}d_t^2\},\\
    \end{split}
\end{align*}
where $j, \ell$ are time indexes, $d_t=t_j-t_\ell$, and $\delta_{j\ell}=1$ if $j=\ell$, otherwise $\delta_{j\ell}=0$.

The covariance function between the time $t_j$ of process $a$ and $t_\ell$ of process $b$, denoted as $C_{ab}^{Y}(t_j,t_\ell)$, can be expressed as,
\begin{align*}
    \begin{split}
      C_{ab}^{Y}(t_j,t_\ell)&=C_{ab}^{\xi}(t_j,t_\ell), \ a \ne b, \\
      C_{ab}^{\xi}(t_j,t_\ell)&=v_{a0}v_{b0}\frac{(2\pi)^{\frac{1}{2}}}{\sqrt{|B_{a0}+B_{b0}|}}\text{exp}\{-\frac{1}{2}B_{ab} d_t^2\},
    \end{split}
\end{align*}
where $B_{ab}= \frac{B_{a0}B_{b0}}{B_{a0}+B_{b0}}$.

Note that the original paper \citep{boyle2005multiple} provides derivation results under more generalized cases. Let $Q$ denote the dimension of the input variable $t_j$, and $M$ the number of shared base process $\tau_0(t)$. In our proposed approach, $Q=1$ and $M=1$. In \cite{boyle2005multiple}, $Q$, $M$ can be arbitrary positive integers.

\subsection*{A.3 Identifiability issue of the model}
Similar to the classical factor analysis, we have identifiability issue caused by scaling and label switching. Differently, we do not have rotational invariance, due to the assumption of correlated factor trajectories (see \cite{zhao2016bayesian} for a detailed description of identifiability issue caused by rotation, scaling and label switching under the classical factor analysis). Instead, we have a new source of non-identifiability, which is the covariance matrix itself. Below we explain specific issues in our model using mathematical expression and describe our corresponding solutions.

To facilitate illustrating the identifiability issue, we first re-express the proposed model in Section 2.3 as,
\begin{equation}
\begin{split}
\vect({\mathbf{X}_i^{T}})&=\vect({\mathbf{M}_i^{T}})+\mathbf{L}_i^{*} \vect({\mathbf{Y}_{i,\text{obs}}^{T}})+ \vect({\mathbf{E}_i^{T}}), \\
 \vect({\mathbf{Y}_{i}}^{T}) &\sim \text{MVN}(\mathbf{0},\Sigma_{\mathbf{Y}}), \\
\vect({\mathbf{E}_i^{T}}) &\sim \text{MVN}(\mathbf{0},\Sigma_{\mathbf{X}_i}),\\
\end{split}
\end{equation}
where $\vect({\mathbf{X}_i^{T}})$, $\vect({\mathbf{M}_i^{T}})$, and $\vect({\mathbf{E}_i^{T}})$ are vectorized from matrices $\mathbf{X}_{i}^{T}$, $\mathbf{M}_{i}^{T}$, and $\mathbf{E}_{i}^{T}$, respectively. To make the above equations hold, $\mathbf{L}_i^{*}$ and $\Sigma_{\mathbf{X}_i}$ are constructed using components of $\mathbf{L}$ and $\boldsymbol{\phi}$, respectively. Specific forms are available in Supplementary Section A.1.


The distribution of $\vect({\mathbf{X}_i^{T}})$ after integrating out $\vect({\mathbf{Y}_{i,\text{obs}}^{T}})$, is 
\begin{equation}
\label{marginal_distribution}
\begin{split}
\vect({\mathbf{X}_i^{T}}) \ | \ \vect({\mathbf{M}_i^{T}}), \mathbf{L}_i^{*}, \Sigma_{\mathbf{X}_i}, \Sigma_{\mathbf{Y}} \sim \text{MVN}(\vect({\mathbf{M}_i^{T}}), \mathbf{L}_i^{*}\Sigma_{\mathbf{Y}_{i,\text{obs}}}(\mathbf{L}_i^{*})^{T}+\Sigma_{\mathbf{X}_i}),
\end{split}
\end{equation}
where $\Sigma_{\mathbf{Y}_{i,\text{obs}}}$ is a sub-matrix of $\Sigma_{\mathbf{Y}}$ that characterizes the covariance structure for $\vect({\mathbf{Y}_{i,\text{obs}}^{T}})$. \\

First, there is identifiability issue with the covariance matrix $\Sigma_{\mathbf{Y}_{i,\text{obs}}}$. This issue arises from the invariance of the covariance $\mathbf{L}_i^{*}\Sigma_{\mathbf{Y}_{i,\text{obs}}}(\mathbf{L}_i^{*})^{T}+\Sigma_{\mathbf{X}_i}$ in Equation \ref{marginal_distribution}. The uniqueness of $\Sigma_{\mathbf{X}_i}$ has been ensured in previous research \citep{ledermann1937rank, bekker1997generic, conti2014bayesian, papastamoulis2022identifiability}; given its identifiability, we are concerned with identifiability of $\Sigma_{\mathbf{Y}_{i,\text{obs}}}$. Non-identifiability is present because for any non-singular transformation matrix $\mathbf{D} \in \mathbb{R}^{kq_i \times kq_i}$, the expression $\mathbf{L}_i^{*}\Sigma_{\mathbf{Y}_{i,\text{obs}}}(\mathbf{L}_i^{*})^{T}+\Sigma_{\mathbf{X}_i}$ are equal under these two sets of estimators: the first estimator is $\{\widehat{\mathbf{L}_i^{*}}, \widehat{\vect({\mathbf{Y}_{i,\text{obs}}^{T}})}, \widehat{\Sigma}_{\mathbf{Y}_{i,\text{obs}}}\}$ and the second estimator is $\{\widehat{\mathbf{L}_i^{*}}\mathbf{D}, \mathbf{D}^{-1}\widehat{\vect({\mathbf{Y}_{i,\text{obs}}^{T}})}, \mathbf{D}^{-1}\widehat{\Sigma}_{\mathbf{Y}_{i,\text{obs}}}(\mathbf{D}^{-1})^{T}\}$. 

To address the issue, we place a constraint on $\Sigma_{\mathbf{Y}}$ that requires its main diagonal element to be $1$.  In other words, the covariance matrix of latent factors $\Sigma_{\mathbf{Y}}$ is forced to be a correlation matrix by assuming the variance of factor to be $1$. This restriction has also been used in \cite{conti2014bayesian}, with multiple purposes: first, it ensures the uniqueness of $\Sigma_{\mathbf{Y}}$; in turn, it helps set the scale of $\vect({\mathbf{Y}_{i,\text{obs}}^{T}})$ and consequently helps set the scale of $\mathbf{L}_i^{*}$. In addition, note that the rotational invariance discussed in \cite{zhao2016bayesian} does not exist here, as $\Sigma_{\mathbf{Y}_{i,\text{obs}}} \neq \mathbf{I}$. 

Second, even under the above constraint, there is still identifiability issue with $\{\mathbf{L}_i^{*}, \vect({\mathbf{Y}_{i,\text{obs}}^{T}})\}$. This is because $\mathbf{D}^{-1}\widehat{\Sigma}_{\mathbf{Y}_{i,\text{obs}}}(\mathbf{D}^{-1})^{T}$ might still equal $\widehat{\Sigma}_{\mathbf{Y}_{i,\text{obs}}}$ when $\mathbf{D}$ is a specific signed matrix (diagonal matrix with diagonal elements being $1$ or $-1$) or permutation matrix (under this case, the identifiability problem is also known as \textit{label switching}), as explained in \cite{papastamoulis2022identifiability}. The specific form of the transformation matrix leading to such invariance depends on the real correlation structure of the data. 

To deal with this unidentifiability caused by signed permutation, a common approach is to post-process samples \citep{zhao2016bayesian}. Here, we use the R package ``factor.switch'' developed in \cite{papastamoulis2022identifiability} to align samples across Gibbs iterations. Alignment of $\{\mathbf{L}^{r}, \mathbf{Y}^{r}\}_{r=1,...,R}$ should be completed before inputting factor scores to GPFDA for estimating DGP parameters $\boldsymbol{\Theta}$ within the MCEM algorithm, and also before the final posterior summary of $\{\mathbf{L},\mathbf{Y}\}$ using the Gibbs-After-MCEM samples. For the latter Gibbs sampler, we ran three chains in parallel; therefore alignment should be firstly carried out within each chain, then across chains. 



\subsection*{A.4 Choosing the Gibbs sample size within the MCEM algorithm}

One challenge with implementing the MCEM algorithm is the choice of the Gibbs sample size. First and foremost, the Gibbs-within-MCEM sample size $R$ determines the computational cost for both the Gibbs sampler and EM. Choosing $R$ too small, while saving computational expense for the Gibbs sampler, results in a less precise approximation of the Q-function, and thus leads to significantly slower convergence of the EM algorithm. In contrast, choosing $R$ too large, while providing an accurate approximation of the Q-function and faster EM convergence, results in the time taken for generating the required Gibbs samples becoming prohibitive. Consequently, the choice of $R$ must trade-off the accuracy of the Q-function with the computational cost of generating Gibbs-within-MCEM samples. We provided a literature review on this topic in Supplementary Section A.4.1 and adapted the approach proposed by \cite{caffo2005ascent} to our model in Supplementary Section A.4.2. This approach automatically determines when to increase $R$ dependent on whether the ascent property of the marginal likelihood under EM is preserved or not \citep{wu1983convergence}.

\subsubsection*{A.4.1 Literature review}
A thorough review of strategies for choosing $R$ can be found in \cite{levine2004automated}. We focus on approaches that automatically adjust the sample size to avoid tedious manual tuning; \cite{neath2013convergence} provided a review of this class of approaches. Briefly, there are primarily two methods with  the main difference between them being the criterion to increase the sample size. The first approach \citep{booth1999maximizing,levine2001implementations,levine2004automated} achieves automatic tuning by monitoring the Monte Carlo error associated with each individual parameter, which is the approximation error incurred when using Monte Carlo samples to approximate the exact expectation. Additional samples are needed if the Monte Carlo error for any of the parameters is deemed too large. However, this approach may be difficult to implement here because our experiments with GPFDA revealed that individual DGP parameters are not identifiable: for the same input data, different runs of GPFDA could return differing individual estimates yet still ensure similar estimates of the covariance matrix (therefore similar marginal likelihoods). 

An alternative, proposed by \cite{caffo2005ascent}, considers increasing the sample size dependent on whether the ascent property of the marginal likelihood under EM is preserved or not \citep{wu1983convergence}. For exact EM, where the expectation can be calculated precisely, the likelihood function is non-decreasing as the algorithm progresses. However, under MCEM, where the E-step is estimated with Monte Carlo samples, it is possible for the likelihood to decrease, due to approximation error. The algorithm by \cite{caffo2005ascent} ensures that the likelihood still increases with a high probability, as the algorithm iterates, by introducing a mechanism for rejection of proposed parameter estimates. Specifically, they use the Q-function as a proxy for the marginal likelihood. An updated $\widehat{\boldsymbol{\Theta}}^{(l)}$ will be accepted only when it increases the  Q-function compared to the previous $\widehat{\boldsymbol{\Theta}}^{(l-1)}$; otherwise, the algorithm will increase the sample size and will propose a new $\widehat{\boldsymbol{\Theta}}^{(l)}$ using the larger sample. 

\subsubsection*{A.4.2 Adaptation of Caffo's approach to our model}
To apply the method of \cite{caffo2005ascent} to our model, we begin by writing out the exact Q-function after the $(l-1)$th iteration of EM. Following Equations~3.5 and 3.8 of the main manuscript,   $Q(\boldsymbol{\Theta},\widehat{\boldsymbol{\Theta}}^{(l-1)}) = \mathbb{E}_{\mathbf{Y}}\left[\text{ln}f(\mathbf{Y}|\boldsymbol{\Theta})\middle|\mathbf{X},\widehat{\boldsymbol{\Theta}}^{(l-1)}\right]$.
Thus, the change in the value of the Q-function after obtaining an updated $\widehat{\boldsymbol{\Theta}}^{(l)}$ compared to the current $\widehat{\boldsymbol{\Theta}}^{(l-1)}$ can be represented as, 
        \begin{align*}
          \Delta Q
          &=Q(\widehat{\boldsymbol{\Theta}}^{(l)},\widehat{\boldsymbol{\Theta}}^{(l-1)})-Q(\widehat{\boldsymbol{\Theta}}^{(l-1)},\widehat{\boldsymbol{\Theta}}^{(l-1)})\\
        &=\mathbb{E}_{\mathbf{Y}}\left[\text{ln}\frac{f(\mathbf{Y}|\widehat{\boldsymbol{\Theta}}^{(l)})}{f(\mathbf{Y}|\widehat{\boldsymbol{\Theta}}^{(l-1)})}\,\middle|\, \mathbf{X},\widehat{\boldsymbol{\Theta}}^{(l-1)}\right]\\
          &=\mathbb{E}_\mathbf{Y}[g(\mathbf{Y})],
        \end{align*}
    where $g(\mathbf{Y})=\text{ln}\frac{f(\mathbf{Y}|\widehat{\boldsymbol{\Theta}}^{(l)})}{f(\mathbf{Y}|\widehat{\boldsymbol{\Theta}}^{(l-1)})}$, and the expectation is with respect to $f(\mathbf{Y}|\mathbf{X},\widehat{\boldsymbol{\Theta}}^{(l-1)})$.
Under MCEM, we approximate this change $\Delta Q$ using the approximate Q-function $\widetilde{Q}$ in Equation~3.7,
        \begin{align*}
        \begin{split}
              \Delta \widetilde{Q}&=\widetilde{Q}(\widehat{\boldsymbol{\Theta}}^{(l)},\widehat{\boldsymbol{\Theta}}^{(l-1)})-\widetilde{Q}(\widehat{\boldsymbol{\Theta}}^{(l-1)},\widehat{\boldsymbol{\Theta}}^{(l-1)})\\
            &=\frac{1}{R}\sum_{r=1}^{R} \text{ln}\frac{f(\mathbf{Y}^{r}|\widehat{\boldsymbol{\Theta}}^{(l)})}{f(\mathbf{Y}^{r}|\widehat{\boldsymbol{\Theta}}^{(l-1)})}\\
            &=\frac{1}{R}\sum_{r=1}^{R}g(\mathbf{Y}^{r}),\qquad\ \mathbf{Y}^{r} \sim f(\mathbf{Y}|\mathbf{X},\widehat{\boldsymbol{\Theta}}^{(l-1)})\\
            &=\overline{g}_{R}.
        \end{split}
        \end{align*}
        
       A generalized version of the Central Limit Theorem (CLT) \citep{jones2004markov} shows that $\overline{g}_{R}$ converges, in distribution, to N($\mathbb{E}_\mathbf{Y}[g(\mathbf{Y})],\frac{\zeta}{R}$) as $R \rightarrow \infty$, where $\zeta = \text{Var}(g(\mathbf{Y}))$ can be estimated using either the sample variance of $g(\mathbf{Y}^{r})$ when the samples $\{\mathbf{Y}^{r}\}_{r=1,...,R}$ are independent, or the batch means approach \citep{lunn2013bugs,guan2019fast} when the samples are dependent (as is the case here, because they are obtained using an MCMC sampler). This implies that, 
        \begin{align*}
        \begin{split}
        &\text{P}\left(\frac{\overline{g}_{R}-\mathbb{E}_\mathbf{Y}[g(\mathbf{Y})]}{\sqrt{\frac{\widehat{\zeta}}{R}}}<Z_{1-\alpha}\right) \approx 1-\alpha; \text{or equivalently,} \\
        &\text{P}\left(\Delta Q>\Delta \widetilde{Q}-\sqrt{\frac{\widehat{\zeta}}{R}}Z_{1-\alpha}\right) \approx 1-\alpha,
        \end{split}
        \end{align*}
        where $\widehat{\zeta}$ is the estimate of $\zeta$, $Z_{1-\alpha}$ is the upper $\alpha$ quantile of the standard normal distribution. $(\Delta \widetilde{Q}-\sqrt{\frac{\zeta}{R}}Z_{1-\alpha})$ is called the ``Lower Bound'' (LB) for $\Delta Q$ \citep{caffo2005ascent} because there is a high chance that $\Delta Q$ is larger than this estimator if we choose $\alpha$ to be small. When LB is positive, it is highly likely that $\Delta Q$ is also positive. The automatic updating rule for the sample size is based on LB. In the $l$th iteration, if LB is positive, then we accept the updated $\widehat{\boldsymbol{\Theta}}^{(l)}$ and keep the current sample size $R$; otherwise, we reject $\widehat{\boldsymbol{\Theta}}^{(l)}$ and continue generating additional samples under $\widehat{\boldsymbol{\Theta}}^{(l-1)}$ for updating again. \cite{caffo2005ascent}  suggests a geometric rate of increase for the sample size, by drawing additional $\frac{R}{m}$ samples for some fixed $m$.

\subsection*{A.5 Specifying the stopping condition}
\label{stopping_condition}
The other challenge when implementing the MCEM algorithm is to specify a stopping criterion. 
We propose to stop the algorithm when the total number of sample size increase exceeds a pre-specified value $W$; by doing so, the number of Gibbs samples and samples input to GPFDA is bounded, therefore ensuring the MCEM algorithm can return results within reasonable time. 

Finally, post-processing Gibbs samples $\{\mathbf{Y}^{r}\}_{r=1,...,R}$ (including burn-in and thinning) before inputting them to GPFDA can help drastically reduce the computational cost of GPFDA while still preserving most information in the samples. 

\subsection*{A.6 Metrics used for assessing the performance of models}
We assessed the performance of predicting gene expressions using three metrics: mean absolute error ($\text{MAE}_{\mathbf{X}}$), mean width of the $95\%$ predictive interval ($\text{MWI}_{\mathbf{X}}$), and proportion of genes within the $95\%$ predictive interval ($\text{PWI}_{\mathbf{X}}$). 
\begin{equation*}
\begin{split}
\text{MAE}_{\mathbf{X}} & = \frac{\sum_{i=1}^{n}\sum_{j=q+1}^{q+u}\sum_{g=1}^{p} |x_{ijg}^{0.5}-x_{ijg}^{true}|}{nu_{2}p}, \\
\text{MWI}_{\mathbf{X}} &= \frac{\sum_{i=1}^{n}\sum_{j=q+1}^{q+u}\sum_{g=1}^{p} |x_{ijg}^{0.975}-x_{ijg}^{0.025}|}{nu_{2}p}, \\
\text{PWI}_{\mathbf{X}} &= \frac{\sum_{i=1}^{n}\sum_{j=q+1}^{q+u}\sum_{g=1}^{p} I(x_{ijg}^{0.025} < x_{ijg}^{true}<x_{ijg}^{0.975})}{nu_{2}p},
\end{split}
\end{equation*}
where $n$, $p$, $q$, $u$ denotes the number of subjects, biomarkers, training time points and test time points, respectively; $x_{ijg}^{true}$ denotes the true value of $x_{ijg}$, and $x_{ijg}^{0.025}, x_{ijg}^{0.5}, x_{ijg}^{0.975}$ denotes $2.5\%$, $50\%$, and $97.5\%$ quantiles of the posterior samples, respectively; and $I(x_{ijg}^{0.025} < x_{ijg}^{true}<x_{ijg}^{0.975})=1$ if $x_{ijg}^{true}$ is within the predictive interval, otherwise it is $0$.

Similarly, we assessed the overall performance of estimating factor trajectories using mean absolute error $\text{MAE}_{\mathbf{Y}}$,
\begin{equation*}
\text{MAE}_{\mathbf{Y}} = \frac{\sum_{i=1}^{n}\sum_{j=1}^{q}\sum_{a=1}^{k}|y_{ija}^{0.5}-y_{ija}^{true}|}{nqk},
\end{equation*}
where $n$, $k$, $q$ denotes the number of subjects, factors, and training time points, respectively; $y_{ija}^{true}$ is the true value of $y_{ija}$, and $y_{ija}^{0.5}$ is the $50\%$ quantile of the posterior samples.

\section*{B. Implementation details of the proposed approach}
\subsection*{B.1 Obtaining good initial values}
To provide good initial values for MCEM, we implemented a two-step approach using available software. First, point estimates of latent factor scores $y_{ija}$ were obtained using BFRM. We centered the gene expression within each individual before inputting it into BFRM: the input data $\mathbf{X}^{c}$ consisted of the centered $x_{ijg}^{c}$, $x_{ijg}^{c}=x_{ijg}-\frac{\sum_{j=1}^{q_i} x_{ijg}}{q_i}$. We did so because BFRM assumes independent data, therefore the intercept is  specific to only a gene and not to a subject; in other words, it cannot estimate subject-gene mean $\mu_{ig}$. Second, initial values of GP parameters were obtained by inputting $\mathbf{Y}$ to GPFDA, with either a DGP or IGP specification.

Note that occasionally, the software BFRM may return an extremely bad initial (i.e., far from the truth), which may consequently result in a local optimum during MCEM rather than a global optimum. To ensure a good estimate of DGP hyperparameters is identified, we suggest users fit the model with multiple initials from BFRM and compare prediction performance for model selection. 

\subsection*{B.2 Specification of hyperparameters and other tuning parameters}
We specified hyperparameters and other tuning parameters as follows; these specifications were used for both the simulation study and the real-data application. 

Specifically, we chose $c_1=d_1=c_2=d_2=c_3=d_3=10^{-2}$ for an uninformative Inverse-Gamma prior distribution for the variance terms. We also set $c_0=10^{-1} \cdot p$ and $d_0 = (1-10^{-1}) \cdot p$ to obtain a sparsity-inducing prior on $Z_{ga}$, as the prior expectation of $\pi_a$ under this specification is $\mathbb{E}[\pi_a] = \frac{c_0}{c_0 + d_0} = 10^{-1}$, implying that only $10\%$ of genes are expected to be involved in each pathway. For the pre-specified values in the MCEM algorithm, we set the rate of sample size increase to $m=2$ and the maximum number of increasing the sample size to $W=5$.

\clearpage

\subsection*{B.3. Summary of the Gibbs-after-MCEM algorithm}
\begin{algorithm}[hbt!]
\caption{Gibbs-after-MCEM Algorithm for Posterior Summary.} \label{posterior_summary_for_other_variables}
\SetKwInOut{KwIn}{Input}
\SetKwInOut{KwOut}{Output}

\KwIn{
\begin{itemize}
\item Observed gene expression $\mathbf{X}_i$ and time points $\mathbf{t}_i$, $i = 1,...,n$ for $n$ people.
\item Prediction time points $\mathbf{t}^{\text{new}}$.
\item $\widehat{\boldsymbol{\Theta}}^{\text{MLE}}$.
\item Number of parallel chains $C$. 
\end{itemize}
}

\KwOut{
\begin{itemize}
\item Posterior samples of $\boldsymbol{\Omega}$, including the pathway expression $\mathbf{Y}_i, i=1,...,n$ and gene-pathway relationship $\mathbf{L}$.
\item Posterior samples of predicted gene expression $\mathbf{X}_i^{\text{new}}, i=1,...,n$ at $\mathbf{t}^{\text{new}}$.
\end{itemize}
}
\textbf{Step 1: Initialization Step} \\
\begin{itemize}
\item Construct $\widehat{\Sigma}_\mathbf{Y}^{\text{MLE}}$ using $\widehat{\boldsymbol{\Theta}}^{\text{MLE}}$,  $\mathbf{t} = \bigcup\limits_{i=1}^{n}\mathbf{t}_{i}$ and $\mathbf{t}^{\text{new}}$.
\item Specify the Gibbs sample size $R^{\text{final}}$.
\end{itemize}

\textbf{Step 2: Sample Generation Step} \\
Run $C$ chains in parallel, and under each chain: draw $R^{\text{final}}$ samples of $\boldsymbol{\Omega}$ using the Gibbs sampler $f(\boldsymbol{\Omega}|\mathbf{X}, \widehat{\Sigma}_\mathbf{Y}^{\text{MLE}})$, then $\mathbf{X}_i^{\text{new}}$ from $f(\mathbf{X}_i^{\text{new}}|\boldsymbol{\Omega}, \widehat{\Sigma}_\mathbf{Y}^{\text{MLE}})$. Post-process samples by burn-in and thinning, and denote the remaining sample size as $R^{\text{final}}_{\text{remain}}$.

\textbf{Step 3: Sample Alignment Step} \\
Align post-processed samples $\{\mathbf{L}^{r}, \mathbf{Y}_i^{r}\}_{r=1,...,R^{\text{final}}_{\text{remain}}}$ within each chain, then across chains; both can be achieved using R package ``factor.switch''. 



\end{algorithm}

\clearpage 

\section*{C. Simulation study}

\subsection*{C.1 Additional results under the setting in the manuscript}

This section presents additional results under the setting described in the manuscript, where the number of subjects $n = 17$, the number of biomarkers $p = 100$, the number of training time points $q = 8$, and the number of latent factors $k = 4$.

\begin{figure}[hbt!]
\centering
\includegraphics[width=\textwidth]{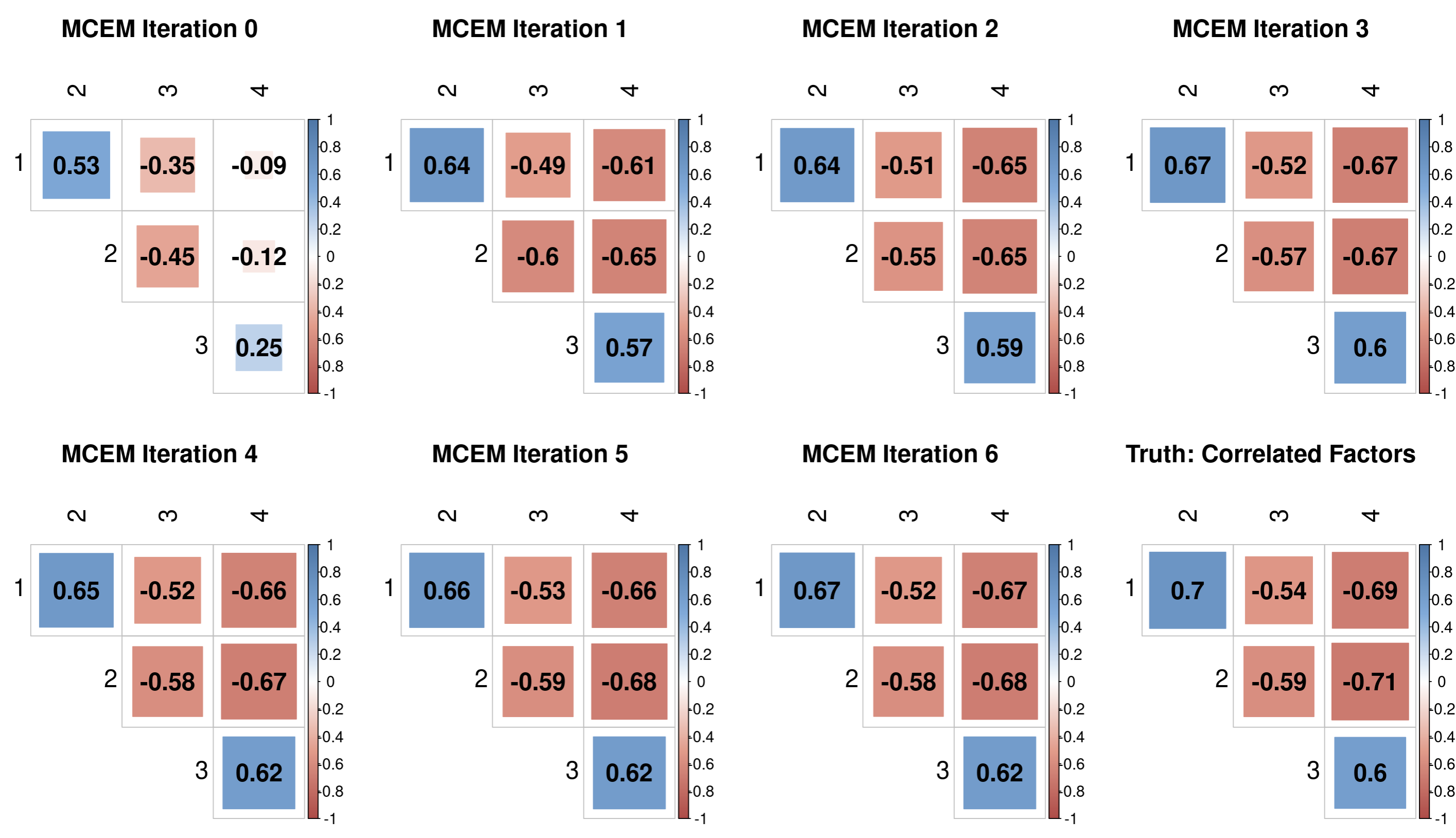}
\caption{An example of cross-correlation matrices at MCEM iterations 0 (initial value) to 6 during the MCEM algorithm using the DGP model: scenario CS, number of latent factors $k$ correctly specified as $4$. The true cross-correlation matrix is also shown.}
\label{sim_cross_correlation_mcem_iteration_process_updated}
\end{figure}

\begin{figure}[hbt!]
\centering
\includegraphics[width=\textwidth]
{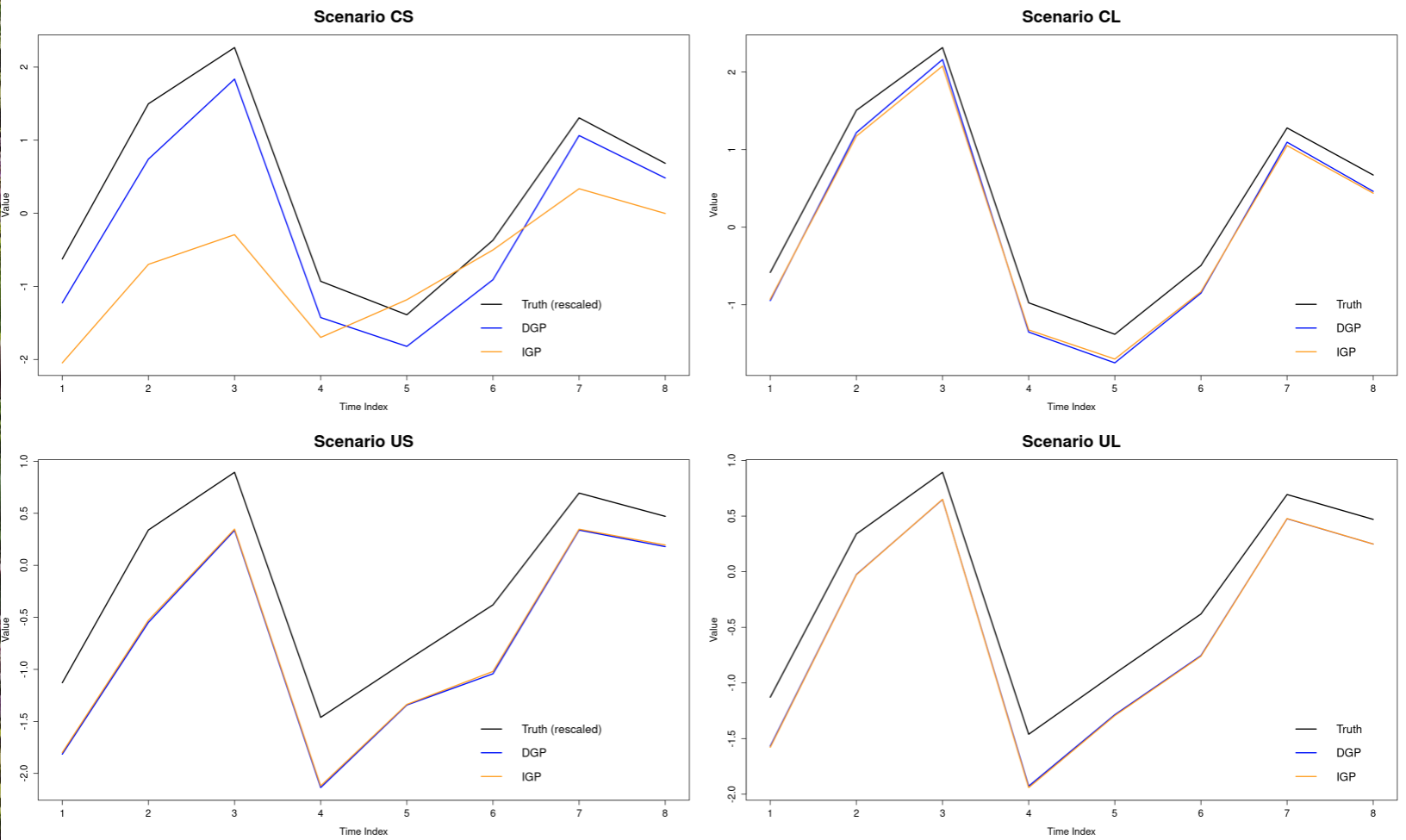}
\caption{Comparison between true and estimated latent factor trajectory for factor 1 of person 1: all scenarios, number of latent factors $k$ correctly specified as $4$. }
\label{recovered_latent_factor_trajectories}
\end{figure}




\begin{figure}[hbt!]
\centering
\includegraphics[width=\textwidth]
{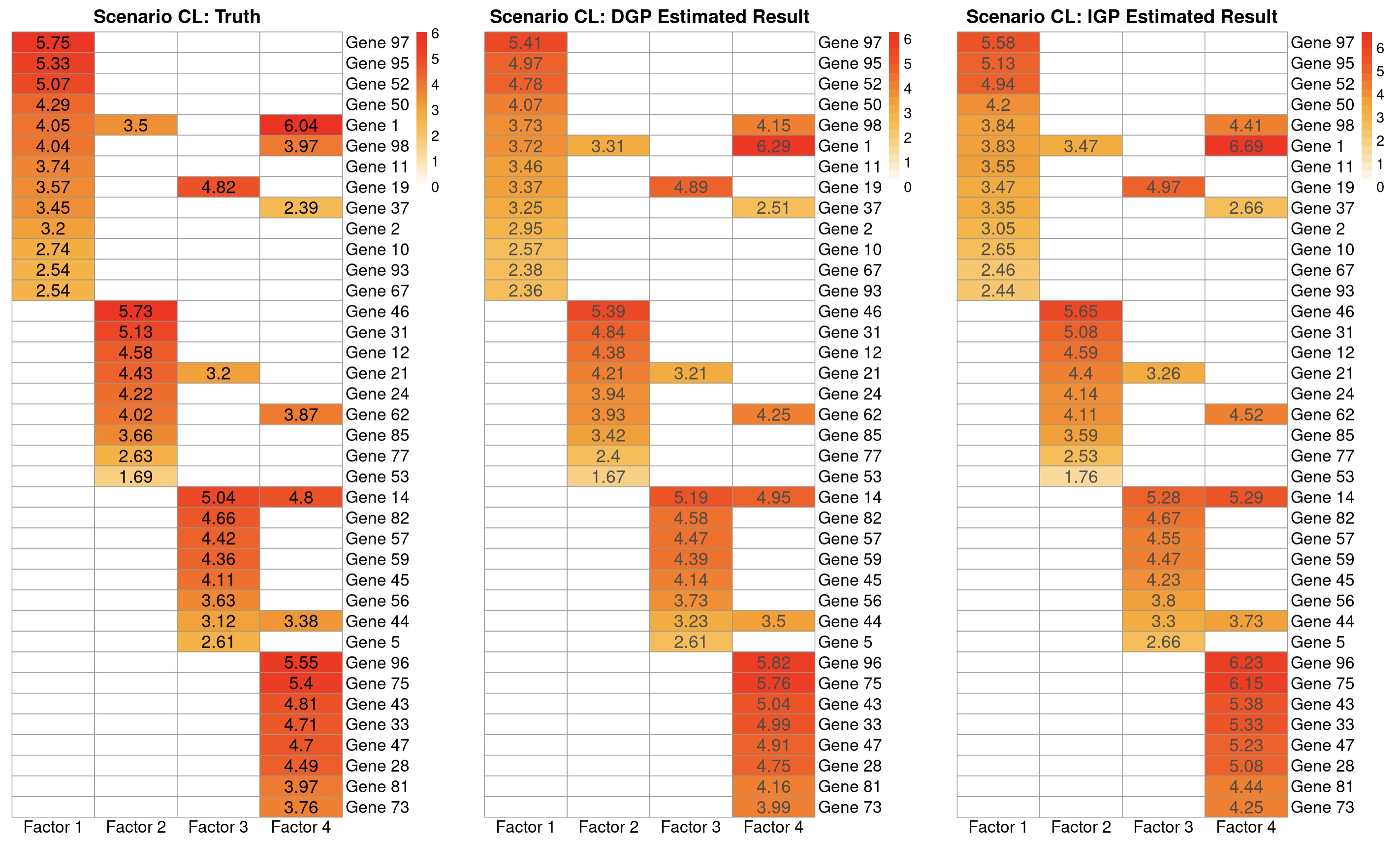}
\caption{Comparison between true and estimated factor loadings: scenario CL, number of latent factors $k$ correctly specified as $4$. Genes displayed in the heatmap of truth (first column) are ordered following two rules: first, genes on factors with smaller indexes are ranked first; second, genes with larger absolute factor loadings are ranked first. Genes displayed in the heatmaps of estimates (second and third columns) follow the ordering of the ground truth to facilitate comparison. }
\end{figure}

\begin{figure}[hbt!]
\centering
\includegraphics[width=\textwidth]
{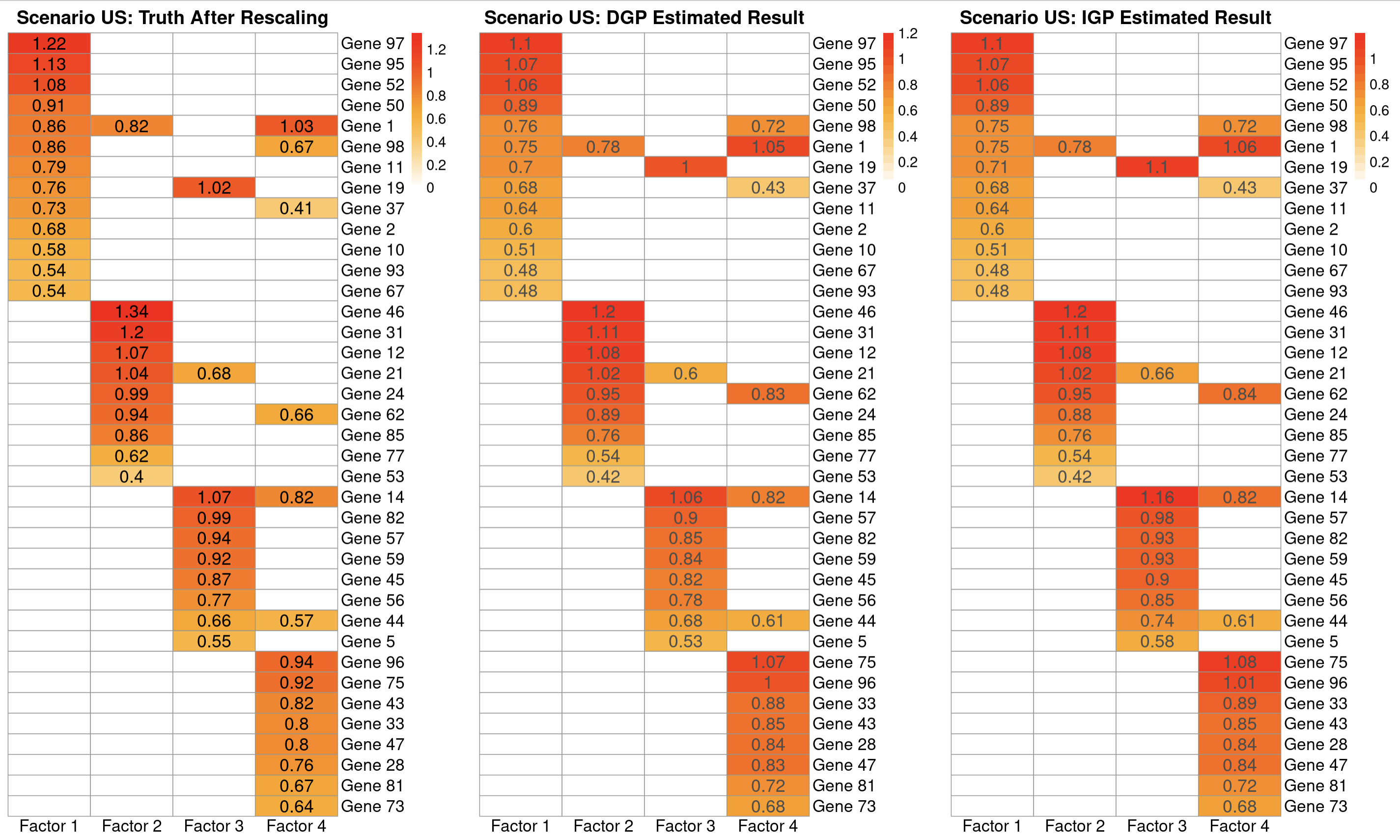}
\caption{Comparison between true and estimated factor loadings: scenario US, number of latent factors $k$ correctly specified as $4$. Genes displayed in the heatmap of truth (first column) are ordered following two rules: first, genes on factors with smaller indexes are ranked first; second, genes with larger absolute factor loadings are ranked first. Genes displayed in the heatmaps of estimates (second and third columns) follow the ordering of the ground truth to facilitate comparison. }
\end{figure}

\begin{figure}[hbt!]
\centering
\includegraphics[width=\textwidth]
{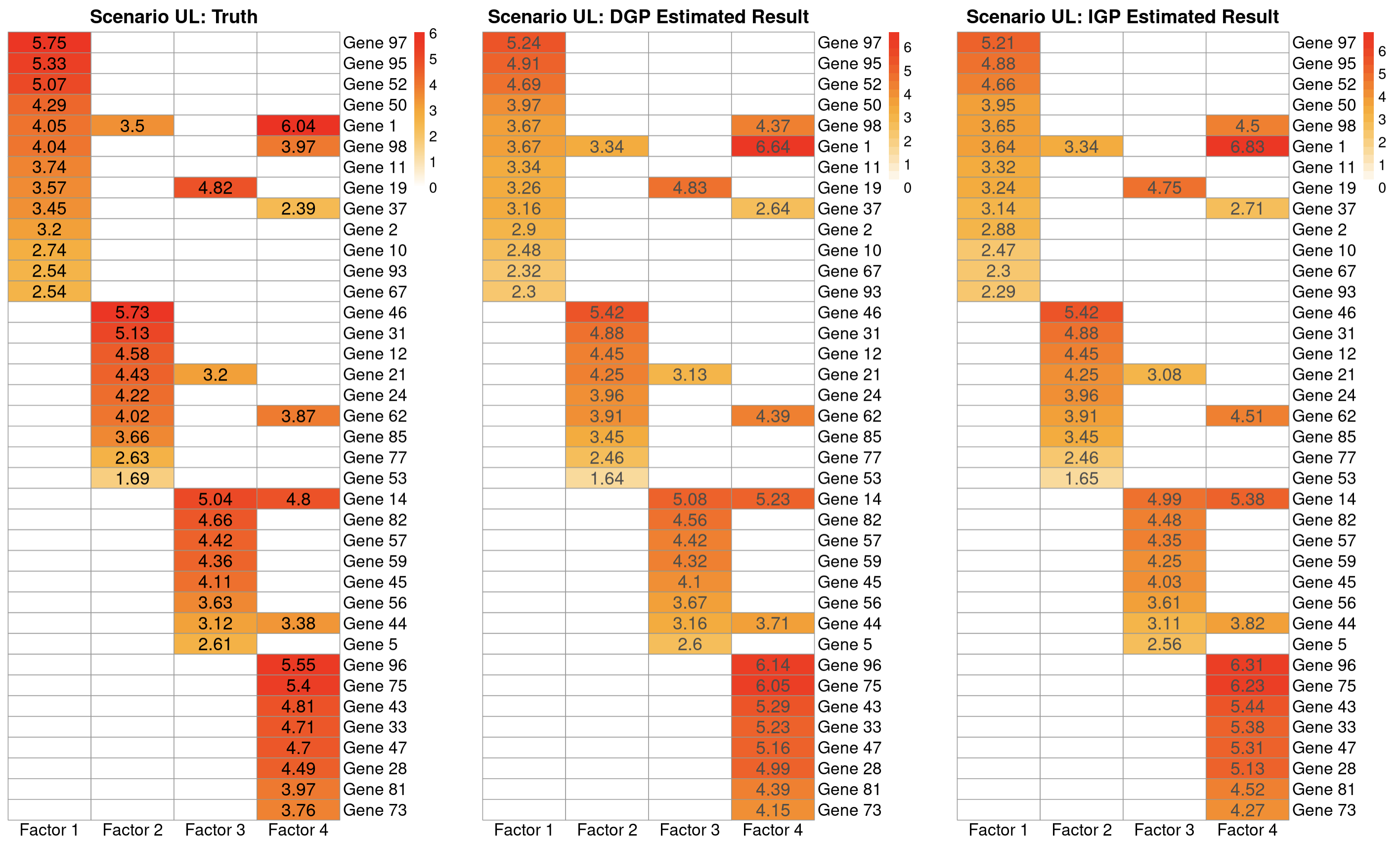}
\caption{Comparison between true and estimated factor loadings: scenario UL, number of latent factors $k$ correctly specified as $4$. Genes displayed in the heatmap of truth (first column) are ordered following two rules: first, genes on factors with smaller indexes are ranked first; second, genes with larger absolute factor loadings are ranked first. Genes displayed in the heatmaps of estimates (second and third columns) follow the ordering of the ground truth to facilitate comparison.}
\label{estimated_factor_loading_matrices_for_scenario_CS_rescale}
\end{figure}

\begin{table}[hbt!]

 \caption{Max Rhat for each type of key variables: all scenarios, number of latent factors $k$ correctly specified as $4$. MCMC iteration number was increased to $100,000$ if Rhat under $10,000$ iterations suggested non-convergence.}
\begin{center}
\resizebox{\textwidth}{!}{\begin{tabular}{l|c|c} 

\hline
\hline
\multicolumn{3}{c}{\textbf{Scenario US}}\\

\hline
\hspace{2cm} \textbf{Model} & DGP  & IGP \\
\textbf{Variable} & MCMC iteration = $10,000$ & MCMC iteration = $10,000$ \\
\hline 
Predicted gene expression & 1.01 & 1.01\\
\hline
Factor loadings & 1.02 & 1.02\\
\hline
Latent factors & 1.01 & 1.01 \\
\hline
\hline
\multicolumn{3}{c}{\textbf{Scenario CS}}\\
\hline
\hspace{2cm} \textbf{Model} & DGP & IGP  \\
\textbf{Variable} & MCMC iteration = $10,000$ & MCMC iteration = $100,000$ ($10,000$) \\
\hline 
Predicted gene expression & 1.01 & 1.04 (1.02)\\
\hline
Factor loadings & 1.01 & 1.09 (4.37)\\
\hline
Latent factors & 1.01 & 1.04 (2.45)\\
\hline
\hline
\multicolumn{3}{c}{\textbf{Scenario UL}}\\
\hline
\hspace{2cm} \textbf{Model} & DGP & IGP \\
\textbf{Variable} & MCMC iteration = $100,000$ ($10,000$) & MCMC iteration = $100,000$ ($10,000$) \\
\hline 
Predicted gene expression & 1.04 (1.01)& 1.04 (1.01)\\
\hline
Factor loadings & 1.04 (1.48)& 1.06 (1.48)\\
\hline
Latent factors & 1.05 (1.40)& 1.06 (1.40)\\
\hline
\hline
\multicolumn{3}{c}{\textbf{Scenario CL}}\\
\hline
\hspace{2cm} \textbf{Model} & DGP & IGP \\
\textbf{Variable} & MCMC iteration = $100,000$ ($10,000$) & MCMC iteration = $100,000$ ($10,000$) \\
\hline 
Predicted gene expression & 1.04 (1.01)& 1.04 (1.01)\\
\hline
Factor loadings & 1.02 (1.29)& 1.06 (1.45)\\
\hline
Latent factors & 1.04 (1.25)& 1.06 (1.37)\\
\hline
\hline

\end{tabular}}
\end{center}
\end{table}

\begin{table}[hbt!]
  \caption{Performance quantification for all scenarios: number of latent factors $k$ correctly specified as $4$ for both DGP and IGP model. {$\text{MAE}_{\mathbf{Y}}$} is short for mean absolute error of estimating latent factors $\mathbf{Y}$.}
  
 \begin{center}
    \begin{tabular}{cc|c|c}
     \hline
    \multicolumn{2}{c|}{Factor Generation Mechanism} & \multicolumn{2}{c} {$\text{MAE}_{\mathbf{Y}}$}\\
    \hline
     Correlation & Variability & DGP &IGP \\
    \hline
     Correlated & Large & 0.15 & 0.16 \\
      & Small & 0.23 & 0.41 \\
     Uncorrelated & Large & 0.16 & 0.17 \\
     & Small & 0.23 & 0.23 \\

    \hline 
    \end{tabular}
   \end{center}
\label{sim_recovering_factor_comparison}
\end{table}

\clearpage

\subsection*{C.2 Further simulation settings under the scenario CS}

This section discusses the performance of our model under more simulation settings. For the ease of illustration, these additional settings were considered under the scenario CS (true factors are correlated and have small variability) only.

\subsubsection*{C.2.1 Mis-specification of the number of latent factors $k$}
We investigated how the mis-specification of $k$ impacts model performance using the scenario CS with $n = 17$, $p = 100$, $q = 8$ as an example. We tried different number of latent factors $k$, ranging from $2$ to $8$. 

Supplementary Table~\ref{compare_results_under_dif_k} shows that correct specification of $k = 4$ leads to the smallest $\text{MAE}_{\mathbf{X}}$ and narrowest $\text{MWI}_{\mathbf{X}}$. 

\begin{table}[hbt!]
   \caption{Prediction performance under different numbers of factors $k$ for scenario CS with $p = 100$ biomarkers. The true number of factors is $k = 4$. $\text{MAE}_{\mathbf{X}}$ and $\text{MWI}_{\mathbf{X}}$ are short for mean absolute error and mean width of the $95\%$ predictive intervals, respectively.}
    \centering
    \begin{tabular}{c|cc}
    \hline
      $k$ & $\text{MAE}_{\mathbf{X}}$ & $\text{MWI}_{\mathbf{X}}$ \\
      \hline
        2 &  0.54 & 2.75 \\
        3 & 0.55 & 2.82\\
        $\mathbf{4}$ & $\mathbf{0.53}$ & $\mathbf{2.66}$\\
        5 & 0.55 & 2.76\\
        6 & 0.54 & 2.78\\
        7 & 0.54 & 2.70\\
        8 & 0.53 & 2.71\\
        \hline
    \end{tabular}
    \label{compare_results_under_dif_k}
\end{table}

Another important metric is factor loading estimates. We have found that our results can always estimate the truth reasonably well regardless of the mis-specification. If $k$ is mis-specified larger than truth, then in practice we have found that the model is able to identify redundant factors (i.e., factors without any non-zero loadings); at the same time, all true factors can be recovered well. If $k$ is mis-specified smaller than truth, then our model can identify a subset of true factors. 

When we specified $k = 5, 6, 7, 8$ (i.e. larger than the true $k = 4$), the additional $1, 2, 3$ or $4$ redundant factors were clearly identified. Taking $k = 5$ as an example for illustration. In the factor loading estimates displayed in Supplementary Figure~\ref{k_5}, there is one factor with no non-zero loadings and the remaining factors are close to the truth. Results under $k = 6, 7, 8$ are similar, as displayed in Supplementary Figures \ref{k_6}, \ref{k_7}, \ref{k_8}, respectively. These observations indicate that our model is able to identify all redundant factors via estimated factor loadings. 

\begin{figure}[hbt!]
\centering
\includegraphics[width=\textwidth]
{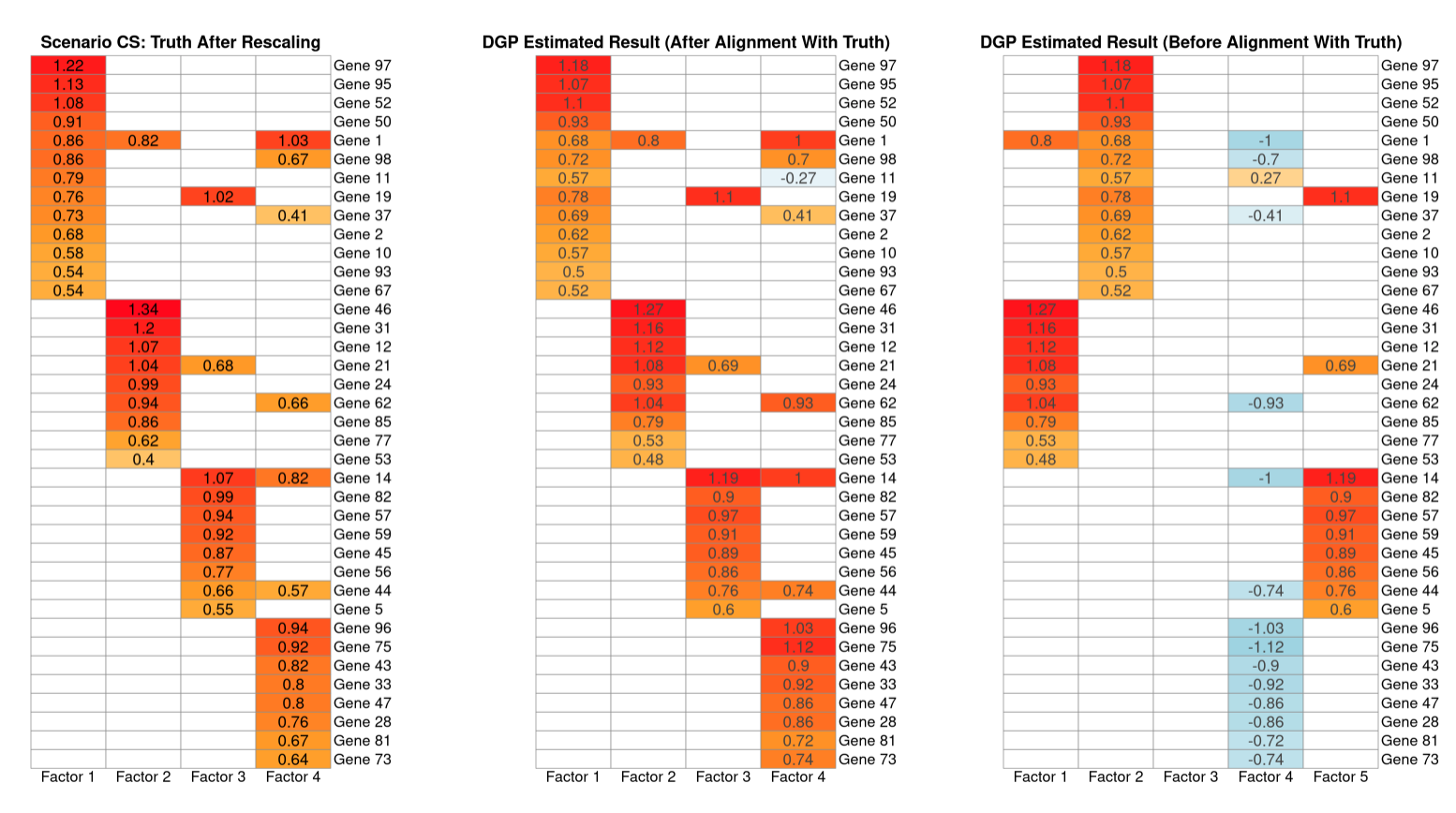}
\caption{Comparison between true and estimated factor loadings using the DGP model: scenario CS, with the number of latent factors $k$ mis-specified as $5$. The second column displays estimates after removal of the redundant factor and alignment with truth. The third column displays raw estimates without alignment with the truth. Alignment is achieved by a signed permutation to facilitate comparison, as explained in \cite{papastamoulis2022identifiability}.}
\label{k_5}
\end{figure}

\clearpage

\begin{figure}[hbt!]
\centering
\includegraphics[width=\textwidth]
{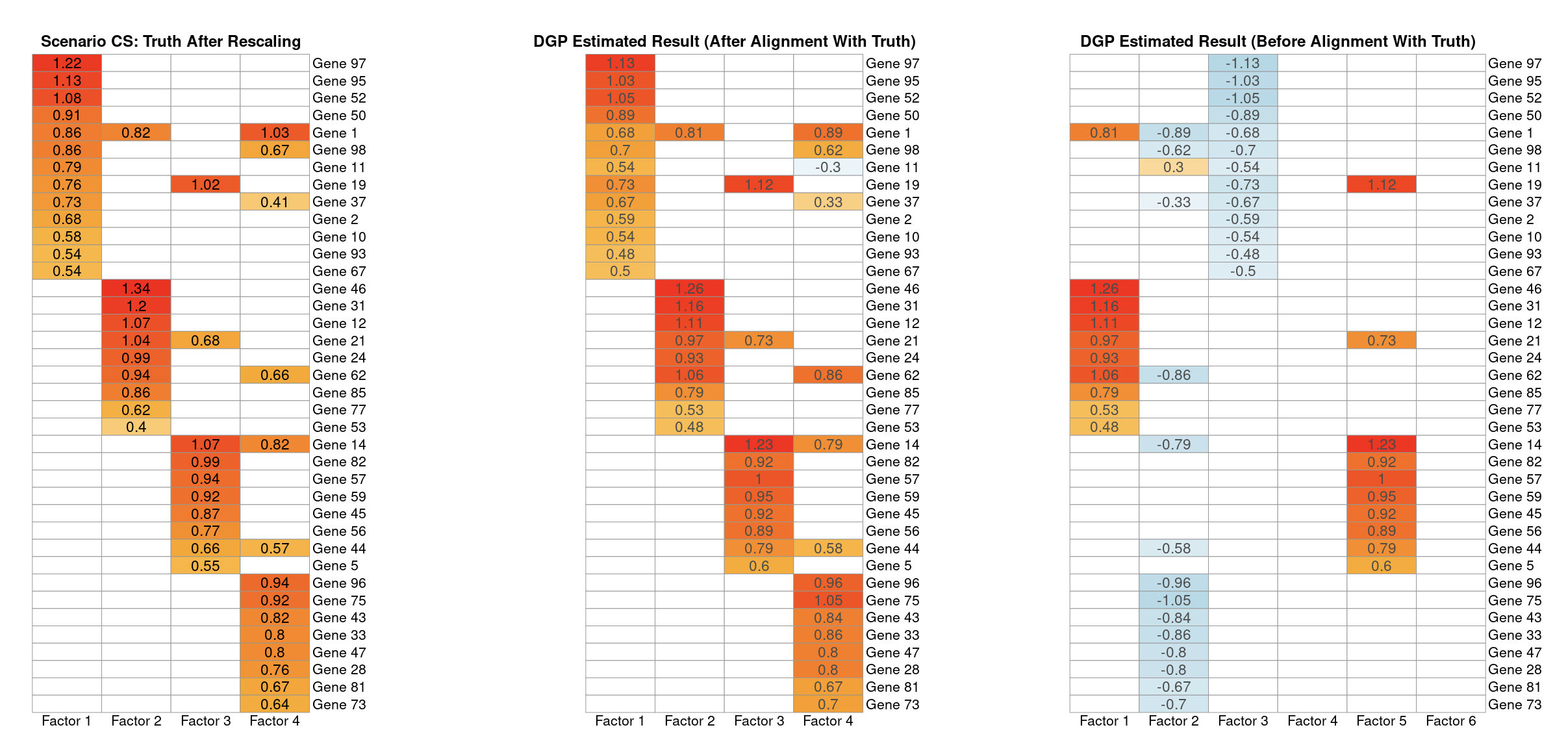}
\caption{Comparison between true and estimated factor loadings using the DGP model: scenario CS, with the number of latent factors $k$ mis-specified as $6$. The second column displays estimates after removal of the redundant factor and alignment with truth. The third column displays raw estimates without alignment with the truth. Alignment is achieved by a signed permutation to facilitate comparison, as explained in \cite{papastamoulis2022identifiability}.}
\label{k_6}
\end{figure}

\clearpage

\begin{figure}[hbt!]
\centering
\includegraphics[width=\textwidth]
{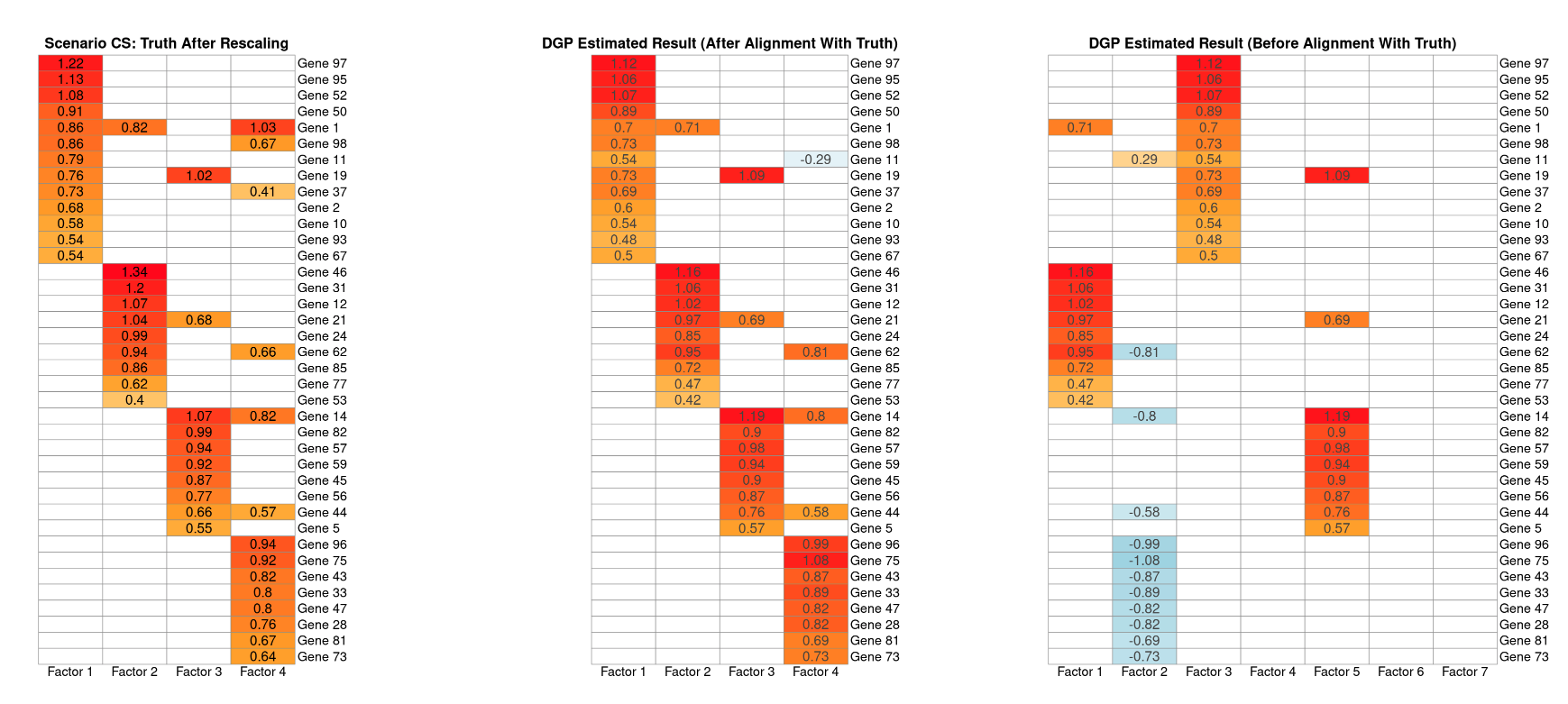}
\caption{Comparison between true and estimated factor loadings using the DGP model: scenario CS, with the number of latent factors $k$ mis-specified as $7$. The second column displays estimates after removal of the redundant factor and alignment with truth. The third column displays raw estimates without alignment with the truth. Alignment is achieved by a signed permutation to facilitate comparison, as explained in \cite{papastamoulis2022identifiability}.}
\label{k_7}
\end{figure}

\clearpage

\begin{figure}[hbt!]
\centering
\includegraphics[width=\textwidth]
{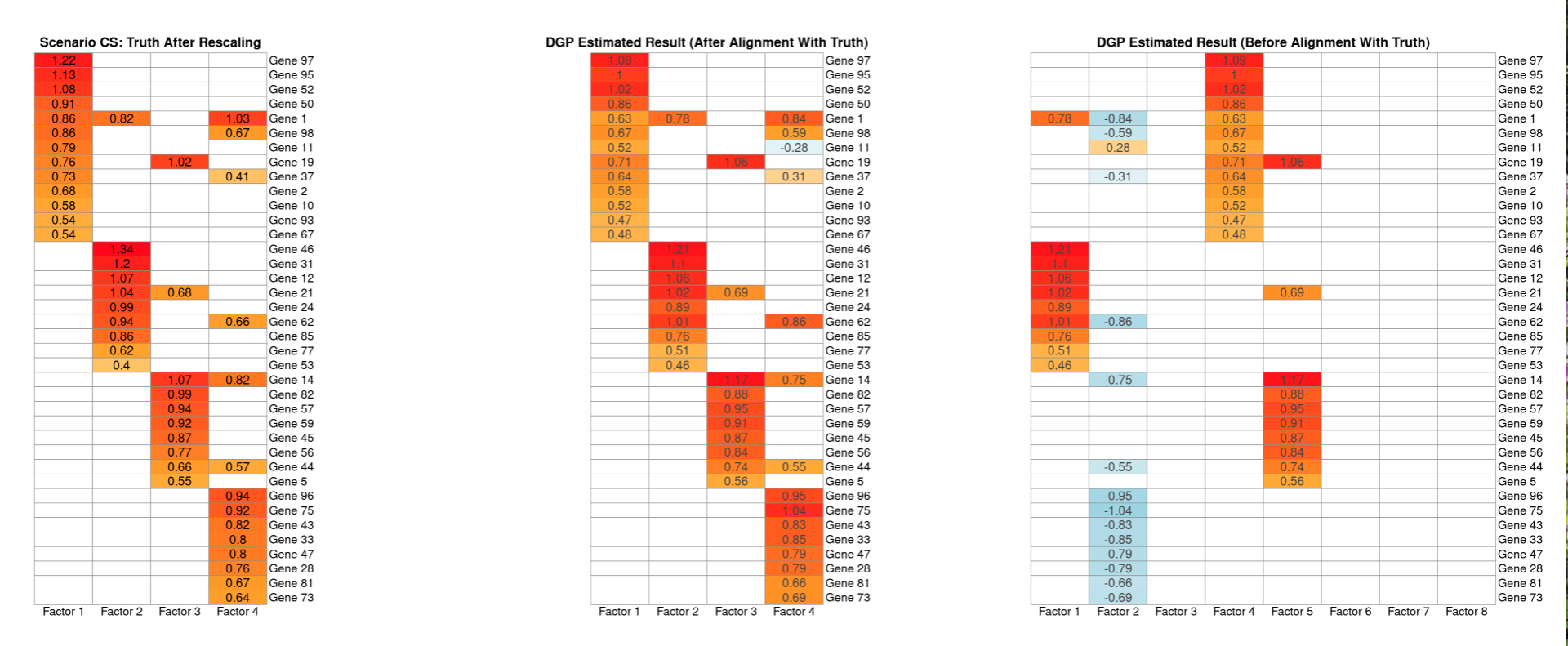}
\caption{Comparison between true and estimated factor loadings using the DGP model: scenario CS, with the number of latent factors $k$ mis-specified as $8$. The second column displays estimates after removal of the redundant factor and alignment with truth. The third column displays raw estimates without alignment with the truth. Alignment is achieved by a signed permutation to facilitate comparison, as explained in \cite{papastamoulis2022identifiability}.}
\label{k_8}
\end{figure}

\begin{figure}[htp]
\centering
\includegraphics[width=\textwidth]
{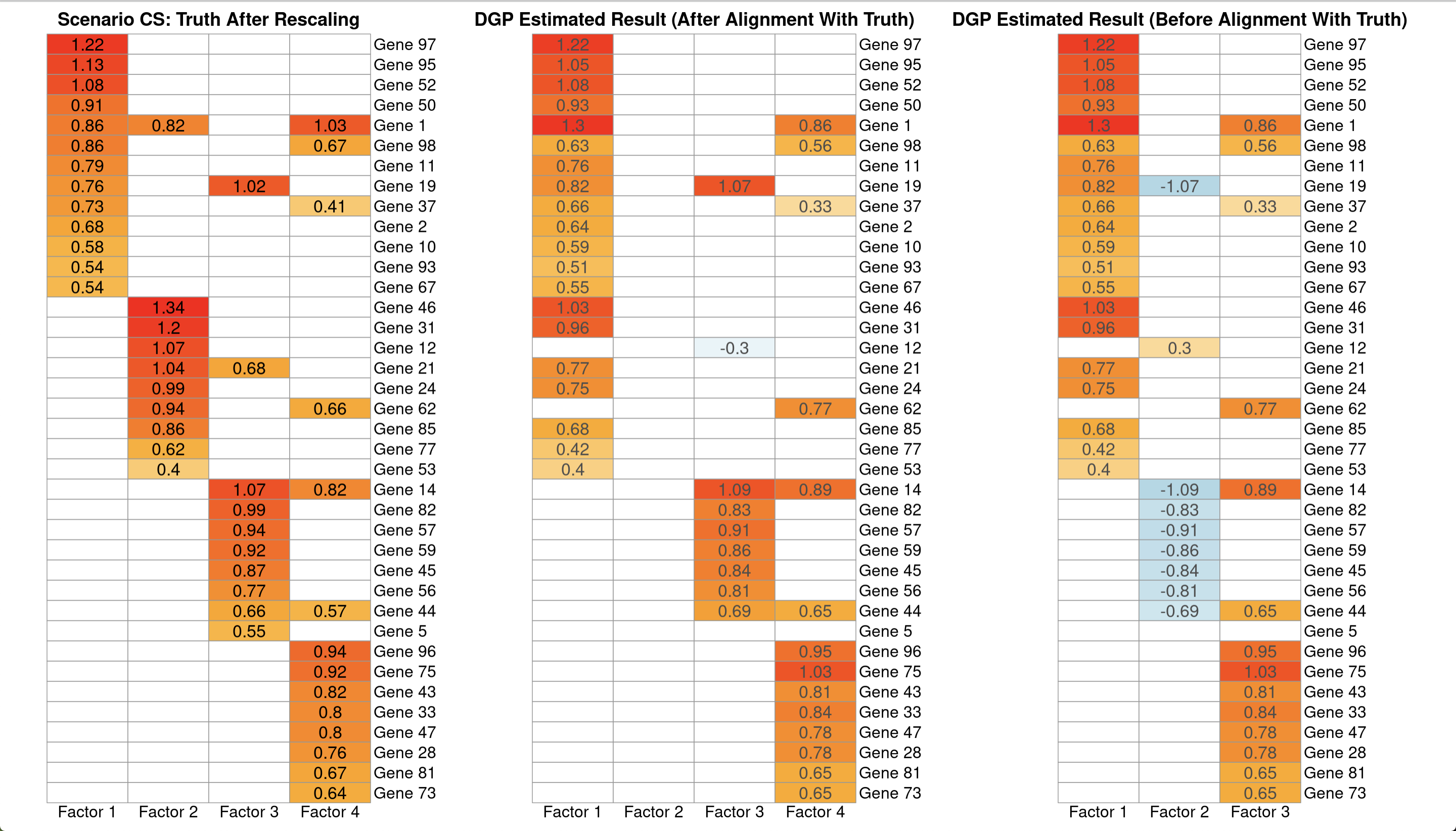}
\caption{Comparison between true and estimated factor loadings using the DGP model: scenario CS, with the number of latent factors $k$ mis-specified as $3$. The second column displays estimates after removal of the redundant factor and alignment with truth. The third column displays raw estimates without alignment with the truth. Alignment is achieved by a signed permutation to facilitate comparison, as explained in \cite{papastamoulis2022identifiability}.}
\label{k_3}
\end{figure}

On the other hand, when $k$ is smaller than the truth, Supplementary Figure~\ref{k_3} shows that the genes loaded on the `missing factor' (true factor 2) loaded on other factors (true factor 1) and the remaining factors approximately match the true loadings. 

\clearpage

\subsubsection*{C.2.2 Impact of the number of biomarkers $p$}
We investigated the performance of our approach under a larger number of genes, with $p = 12000$. Other settings are: sample size $n = 17$ and number of training time points $q = 12$. 

\begin{figure}[htp]
\centering
\includegraphics[width=\textwidth]
{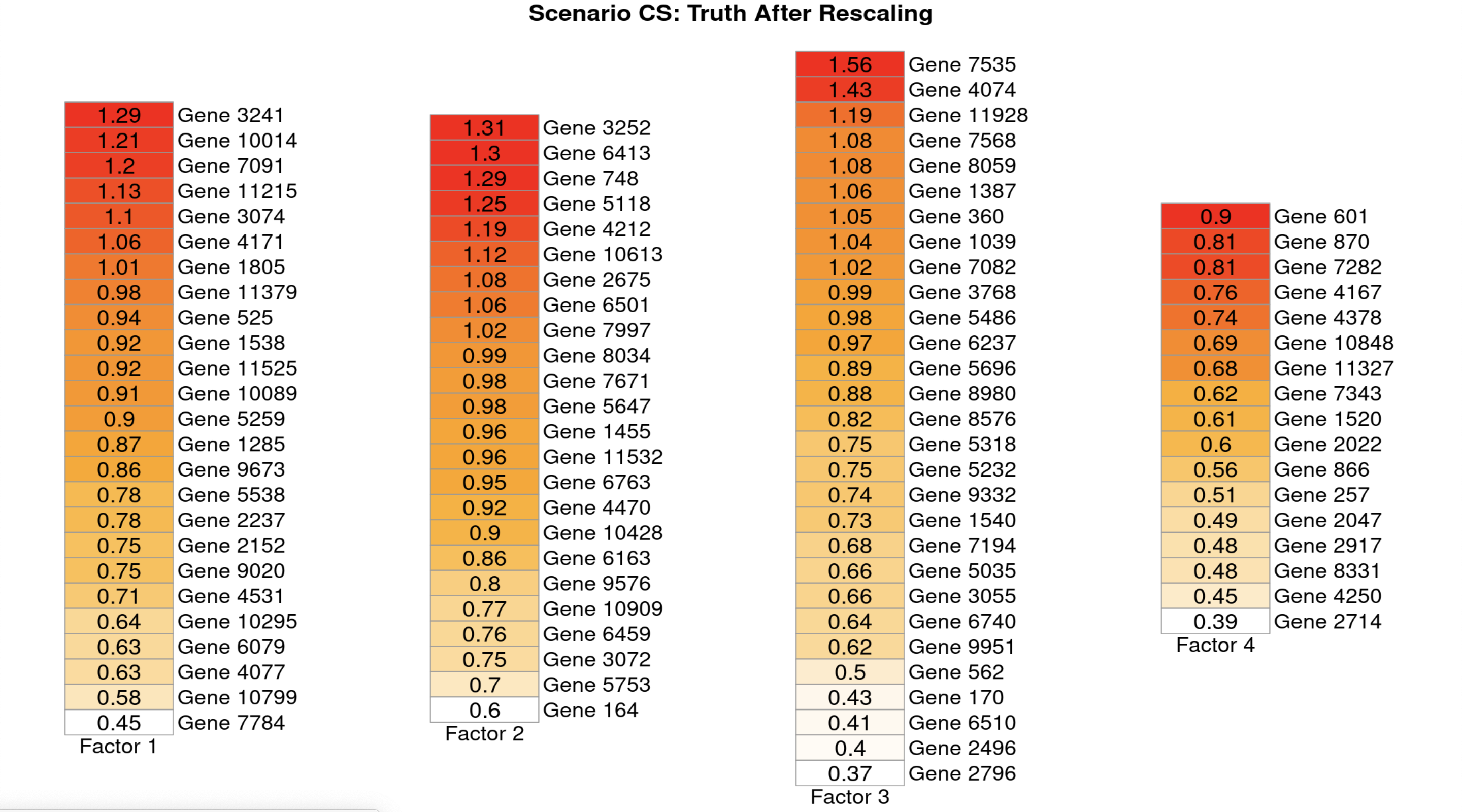}
\caption{True factor loadings: scenario CS under biomarker number $p = 12,000$ and the true number of latent factors $k = 4$. Only non-zero loadings are displayed for each factor.}
\label{cs_large_p_truth}
\end{figure}

\begin{figure}[htp]
\centering
\includegraphics[width=\textwidth]
{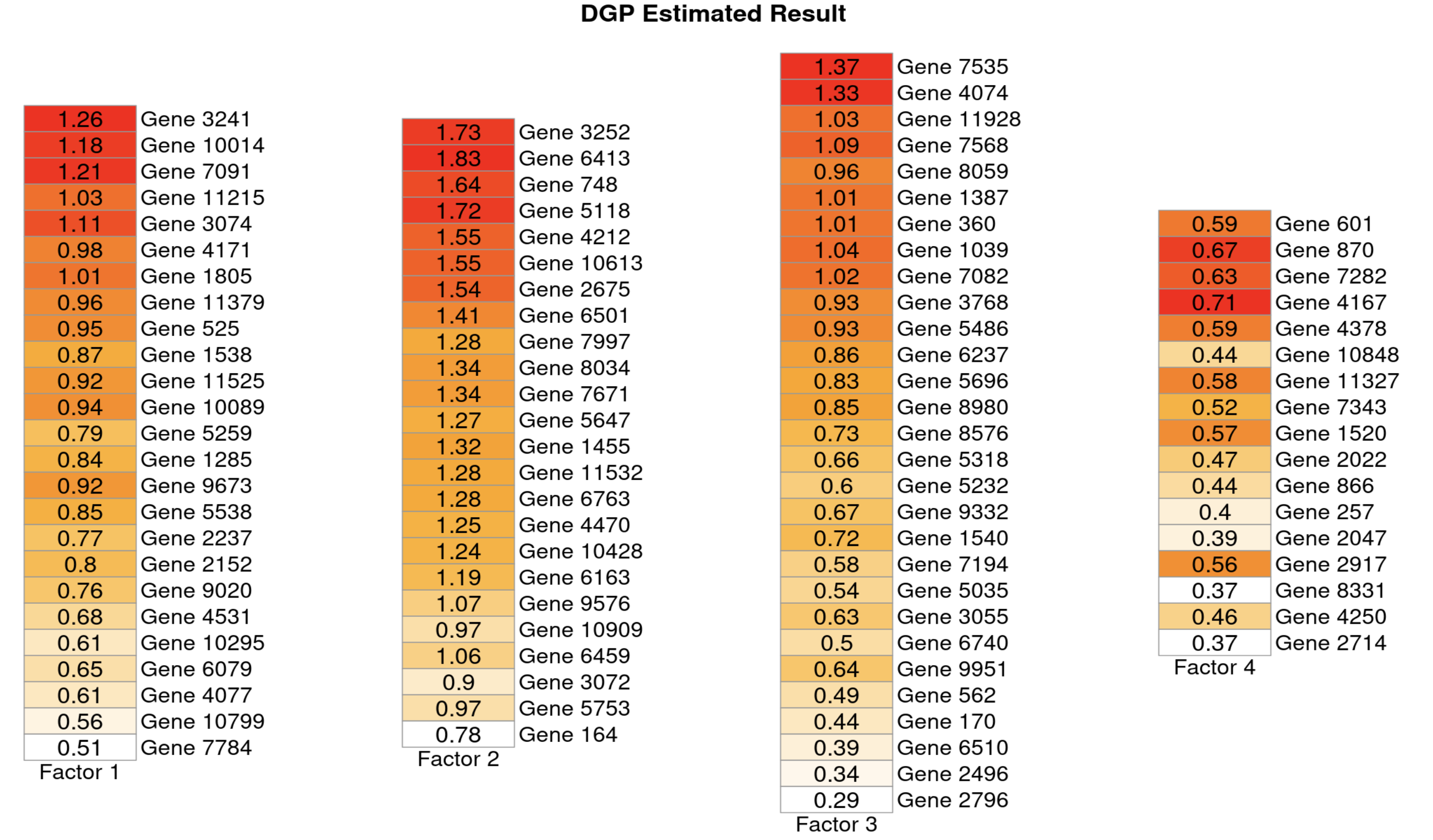}
\caption{Estimated factor loadings using the DGP model: scenario CS under biomarker number $p = 12,000$ and the number of latent factors is correctly specified as $k = 4$. Estimated results are displayed following the gene ordering of the ground truth in Figure~\ref{cs_large_p_truth}.}
\label{cs_large_p_dgp}
\end{figure}

\begin{figure}[htp]
\centering
\includegraphics[width=\textwidth]
{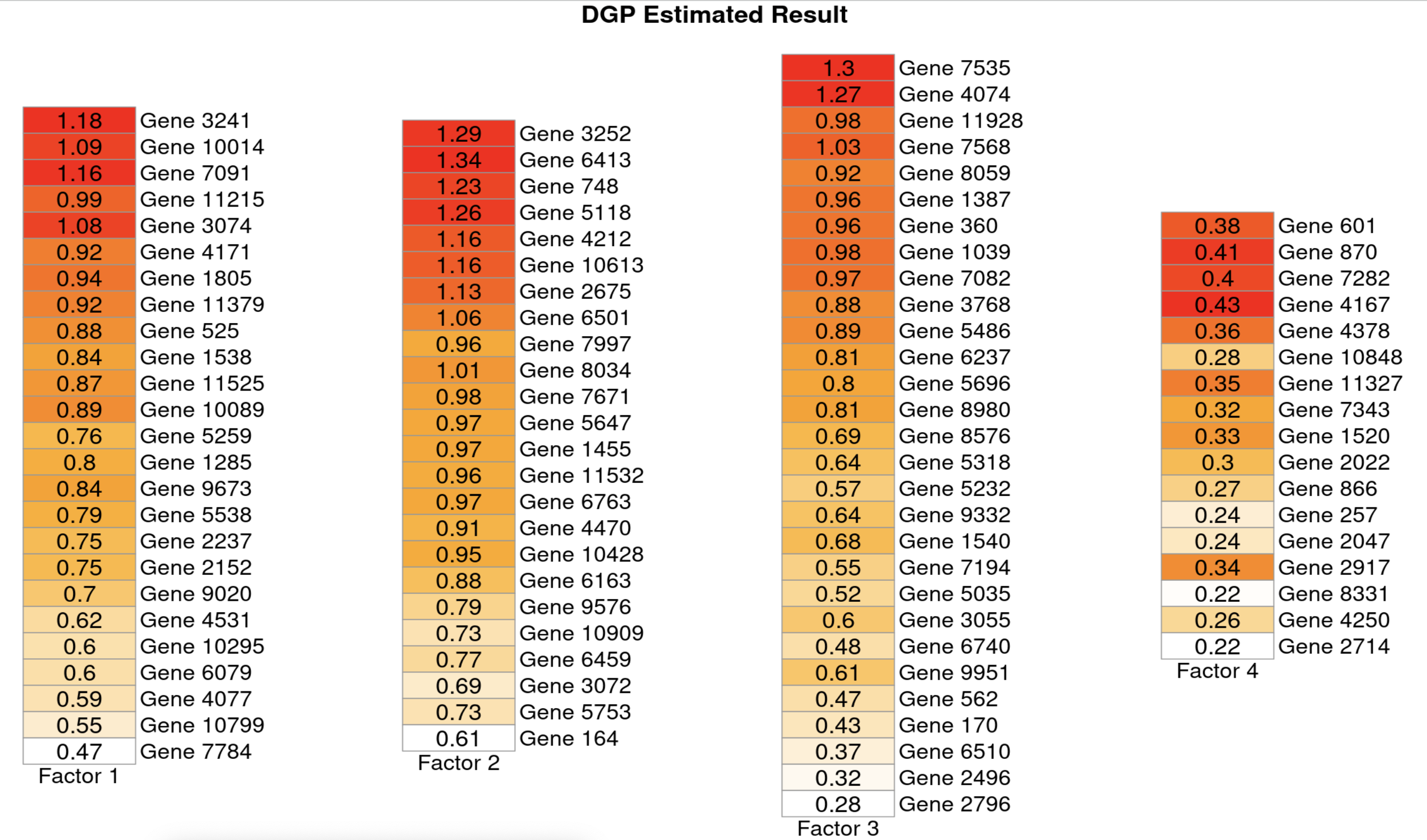}
\caption{Estimated factor loadings using the DGP model: scenario CS under biomarker number $p = 12,000$ and the number of latent factors is mis-specified as $k = 6$. Estimated results are displayed following the gene ordering of the ground truth in Figure~\ref{cs_large_p_truth}.}
\label{cs_large_p_dgp_k_6}
\end{figure}

True factor loadings are displayed in Figure~\ref{cs_large_p_truth}, and estimated results under correct specification of $k = 4$ and incorrect specification $k = 6$ are displayed in Figure~\ref{cs_large_p_dgp} and Figure~\ref{cs_large_p_dgp_k_6}, respectively. Our approach is always able to recover the true factor loadings reasonably well; and when the specified number is unnecessarily large ($k = 6$), our model can identify redundant factors (i.e., corresponding to $2$ columns with all loadings as zero).

Note that the display of results under $p = 12,000$ is different from that under $p = 100$, as the previous way (under $p=100$) of displaying results would lead to small texts hard to read. Therefore, we changed the display format: for the truth, we only display non-zero loadings for each factor; then we followed the referee's suggestion to display the estimated result following the gene ordering of the ground truth. Redundant factors under mis-specifiction of $k = 6$ are not displayed, as corresponding columns do not have non-zero loadings.

\clearpage

\subsubsection*{C.2.3 Impact of the sample size $n$}
We investigated the impact of the number of subjects $n$ on the result. We fix $p = 100$, $q = 8$ and $k = 4$, while we vary $n = 10, 17 (\text{the size in the real data}), 20, 40$. 

We find that as the sample size $n$ increases, the performance of our model will improve. Supplementary Table~\ref{compare_results_under_dif_n} shows comparison in biomarker prediction.

\begin{table}[hbt!]
   \caption{Prediction performance under different numbers of subjects $n$ for scenario CS with $p = 100$ biomarkers. The true number of factors is $k = 4$. $\text{MAE}_{\mathbf{X}}$ and $\text{MWI}_{\mathbf{X}}$ are short for mean absolute error and mean width of the $95\%$ predictive intervals, respectively.}
    \centering
    \begin{tabular}{c|cc}
    \hline
      $n$ & $\text{MAE}_{\mathbf{X}}$ & $\text{MWI}_{\mathbf{X}}$ \\
      \hline
        10 &  0.54 & 2.83  \\
        17 & 0.53 & 2.66 \\
        20 & 0.50 & 2.61 \\ 
        40 & 0.49 & 2.46 \\ 
        \hline
    \end{tabular}
    \label{compare_results_under_dif_n}
\end{table}

\clearpage

\subsubsection*{C.2.4 Impact of the number of time points $q$}
We investigated the impact of the number of time points $q$ on the result. We fix $p = 100$, $n = 17$ and $k = 4$, while we vary $q = 4, 8, 16$. We find that as $q$ increases, the performance of our model will improve. Supplementary Table~\ref{compare_results_under_dif_q} shows comparison in biomarker prediction.

\begin{table}[hbt!]
   \caption{Prediction performance under different numbers of time points $q$ for scenario CS with $p = 100$ biomarkers. The true number of factors is $k = 4$. $\text{MAE}_{\mathbf{X}}$ and $\text{MWI}_{\mathbf{X}}$ are short for mean absolute error and mean width of the $95\%$ predictive intervals, respectively.}
    \centering
    \begin{tabular}{c|cc}
    \hline
      $q$ & $\text{MAE}_{\mathbf{X}}$ & $\text{MWI}_{\mathbf{X}}$ \\
      \hline
        4 &  0.59 & 2.89  \\
        8 & 0.53 & 2.66 \\
        16 & 0.50 & 2.45 \\ 
        25 & 0.45 & 2.21 \\
        \hline
    \end{tabular}
    \label{compare_results_under_dif_q}
\end{table}

\subsubsection*{C.2.5 Impact of the proportion of overlap between factors}
We investigated the impact of greater overlap across factors on the model performance. In literature, it is called `cross-loading' if an observed variable is significantly loaded on more than one latent factor \citep{ximenez2022consequences, li2020effects}. The simulation study in the manuscript demonstrated good performance in recovering the factor loading when the proportion of genes with cross-loading is $21\%$ ($8$ out of the $38$ significant genes are loaded on more than factor).

This additional simulation assesses the performance of our model under a larger cross-loading proportion $60\%$ ($16$ out of the $26$ significant genes are loaded on more than factor). 

\begin{figure}[hbt!]
\centering
\includegraphics[width=\textwidth]
{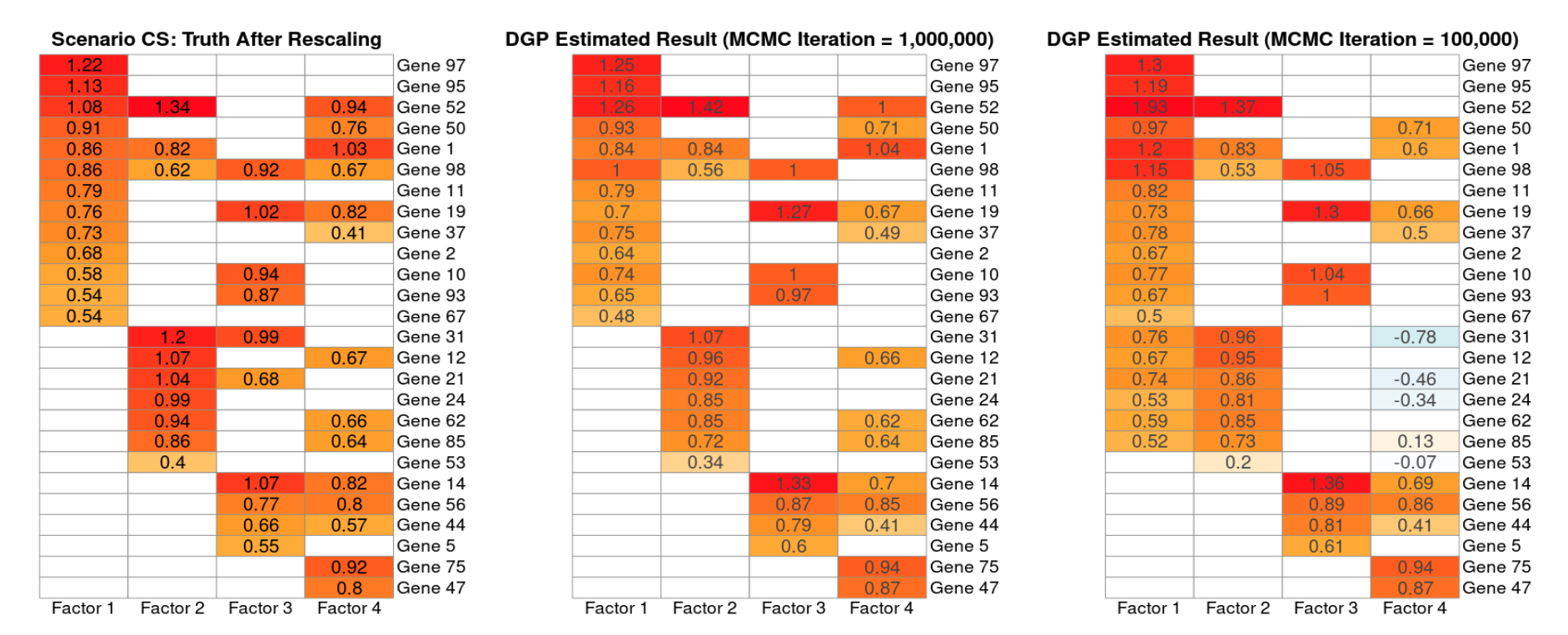}
\caption{Comparison between true and estimated factor loadings using the DGP model: scenario CS with $p = 100$ biomarkers, with the proportion of genes with cross-loadings as $60\%$. The second column displays estimates after 1,000,000 iterations. The third column displays estimates after 100,000 iterations.}
\label{result_greater_overlap}
\end{figure}

Our model continued to perform well at estimating the factor loadings (Supplementary Figure~\ref{result_greater_overlap}), though a longer chain (corresponding to a more expensive computation cost) was required due to the more complex factor structure. In the original setting, only $100,000$ iterations were required while this number was not enough for the new setting with a greater overlap (results displayed in the third column of Supplementary Figure~\ref{result_greater_overlap}); a longer chain such as $1,000,000$ iterations were required (results displayed in the second column of Supplementary Figure~\ref{result_greater_overlap}). 

It is worth highlighting that for each gene that is loaded on more than one factor, the factor with the largest loading can always be identified by our method. Take gene 98 as an example. The true loadings are non-zero on all factors (with the largest loading on factor 3). Although our model incorrectly estimates zero loading on factor 4, it is able to correctly estimate non-zero loading on factor 3.

\clearpage

\section*{D. Real Data}

\subsection*{D.1 Prediction performance under different numbers of factors $k$}

\begin{table}[hbt!]
   \caption{Prediction performance under different factor number $k$ in the H3N2 data application. $\text{MAE}_{\mathbf{X}}$ and $\text{MWI}_{\mathbf{X}}$ are short for mean absolute error and mean width of the $95\%$ predictive intervals, respectively.}
   
    \centering
    \begin{tabular}{c|cc}
    \hline
      Specification of $k$ & $\text{MAE}_{\mathbf{X}}$ & $\text{MWI}_{\mathbf{X}}$ \\
      \hline
        3 & 0.215 & 1.168 \\
        4 & 0.213 & 1.143 \\
        $\mathbf{5}$ & $\mathbf{0.212}$ & $\mathbf{1.118}$ \\
        \hline
    \end{tabular}
    \label{compare_results_under_dif_k_h3n2}
\end{table}

\clearpage 

\subsection*{D.2 Comparison between our model and \cite{chen2011predicting}'s model}
\begin{figure}[hbt!]
\centering
\includegraphics[width=\textwidth]
{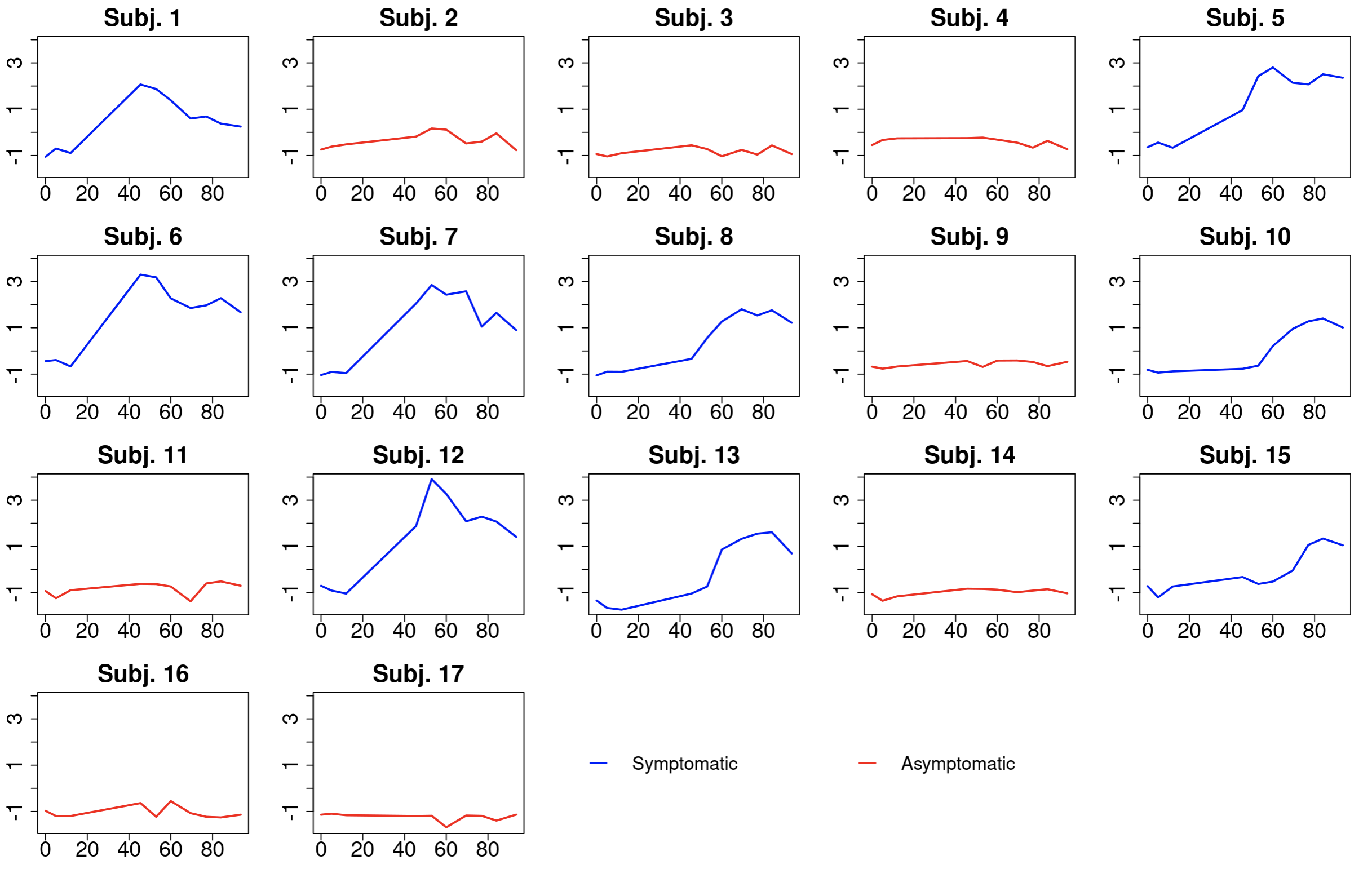}
\caption{Estimated trajectories for factor $1$ in the H3N2 data. Each panel displays a single subject's trajectory, following the same format as Figure 4 in \cite{chen2011predicting}. The horizontal axes correspond to time in hours. This figure appears in color in the electronic version of this article.}
\label{bsfadgp_factor_1_trajectory}
\end{figure}

To facilitate comparison between our factor 1 and their `principal factor', Supplementary Figure \ref{bsfadgp_factor_1_trajectory} displays factor trajectories of all people in the same format as Figure 4 in \cite{chen2011predicting}. For symptomatic people, both factors display an increase after the inoculation (time $0$), then decrease to a level that is higher than before-inoculation; but for asymptomatic people, both factor shapes have little change.  

In addition to the shape similarity, genes significantly loaded on factor $1$ are also largely similar to those loaded on \cite{chen2011predicting}'s principal factor. \cite{chen2011predicting} list the top $50$ genes sorted according to the absolute loading values (from most important to least important); we present this alongside the full list of our top $50$ genes in Supplementary Table \ref{gene_comparison} for comparison. $33$ out of the $50$ genes are the same.

\clearpage 

\setlength{\LTcapwidth}{\linewidth}
\begin{longtable}{c|c|c}

\caption{Top $50$ genes sorted according to absolute factor loadings. Numbers within the parenthesis in the right column represent rank of this gene in the left column, and ``-'' denote this gene is not in the left column. Note that names beginning with ``M97935'' in \cite{chen2011predicting} are actually control sequences rather than genes.} \\


\hline
Row Index & Principal Factor by \cite{chen2011predicting} & Factor 1 by Our Approach\\
\hline
 1& RSAD2 & LAMP3 (13)  \\
2 & IFI44L & RSAD2 (1) \\
3&IFIT1 & IFI44L (2)\\
4&IFI44 & SERPING1 (10)\\
5&HERC5 & SPATS2L (-) \\
6&OAS3 & SIGLEC1 (22)\\
7&MX1 & ISG15 (8) \\
8&ISG15 & IFIT1 (3)\\
9&IFIT3&  IFI44 (4)\\
10&SERPING1 & RTP4 (35) \\
11&IFIT2 & OAS3 (6) \\ 
12&OASL & IFI6 (17) \\ 
13&LAMP3 & CCL2 (-) \\ 
14&IFI27 & IDO1 (-) \\
15&OAS1 & HERC5 (5) \\ 
16&OAS2& MS4A4A (-) \\ 
17&IFI6& IFIT3 (9)\\
18&IFIT5& OAS1 (15) \\ 
19&IFITM3&OAS2 (16)\\
20&XAF1& LY6E (24) \\ 
21&DDX58 & OASL (12)\\ 
22&SIGLEC1& ATF3 (-)\\
23&DDX60& CXCL10 (-) \\ 
24&LY6E & CCL8 (-) \\ 
25&GBP1& XAF1 (20)\\  
26&IFIH1& IFI27 (14)\\ 
26&LOC26010 & SAMD4A (-)\\ 
28&ZCCHC2& MX1 (7) \\ 
29&EIF2AK2& LGALS3BP (-) \\
30&LAP3& C1QB (-) \\ 
31&IFI35 & IFITM3 (19) \\ 
32&IRF7 & LAP3 (30)\\
33&PLSCR1 & IRF7 (32) \\
34&M97935\_MA\_at & ZBP1 (42) \\ 
35&RTP4 &  HERC6 (37)\\ 
36&M97935\_MB\_at & TFEC (-) \\
37&HERC6 & IFI35 (31) \\
38&TNFAIP6 &  MT2A (-) \\
39&PARP12 & SCO2 (41)\\ 
40&M97935\_5\_at & DDX58 (21)\\ 
41&SCO2 & IFIH1 (26)\\ 
42&ZBP1 & IFIT2 (11)\\ 
43&STAT1 & DHX58 (-) \\ 
44&UBE2L6 & TMEM255A (-) \\ 
45& MX2 & TNFAIP6 (38)\\
46&TOR1B & VAMP5 (-) \\
47&M97935\_3\_at & PARP12 (39) \\
48&TNFSF10 & GBP1 (25)\\
49&TRIM22 & TIMM10 (-)\\
50& APOL6 & C1QA (-)\\ 

\hline 
\label{gene_comparison}
\end{longtable}

Despite the similarity between factor $1$ estimated by our BSFA-DGP and the principal factor estimated by \cite{chen2011predicting}, as discussed above, it is noteworthy that the trajectory of factor $1$ is actually more individualized and informative than the principal factor. This can be seen by observing that, for symptomatic people, the shapes of the principal factor are exactly the same after the factor starts changing (Figure 4 in \cite{chen2011predicting}) whereas the shapes of factor $1$ still vary locally for different subjects, as can be seen from Supplementary Figure \ref{bsfadgp_factor_1_trajectory}. 

The difference in the recovered factor trajectories is caused by the different assumptions underlying the two models. \cite{chen2011predicting} assume that the dynamic trajectory of any factor is common to all individuals. Different factor values are observed at the same time point for different people only due to individual-dependent biological time shifts and random noise. Therefore, symptomatic and asymptomatic individuals are distinguished based on the time shift (Figure 4 in \cite{chen2011predicting}). In contrast, we assume that the trajectory of each factor comes from a distribution (i.e., there is no common curve for every individual), therefore allowing for different curve shapes for different people. We make direct use of the diversity of curve shapes to distinguish patients without introducing the ``time shift'' quantity.

An additional advantage of our model compared to that in \cite{chen2011predicting} is its ability to detect potential outliers, due to allowing for subject-specific trajectory. In contrast, Chen et al., assumes a universal factor trajectory and trajectories of different subjects are the result of only shifting this reference trajectory by subject-specific biological time shift along the x-axis. This assumption means extreme values of factor expression (outliers) would be hard to identify, which means additional discoveries may be missed. 

\clearpage 

\subsection*{D.3 Comparison between BSFA-DGP and BSFA-IGP}

\begin{figure}[hbt!]
\centering
\includegraphics[width=\textwidth]
{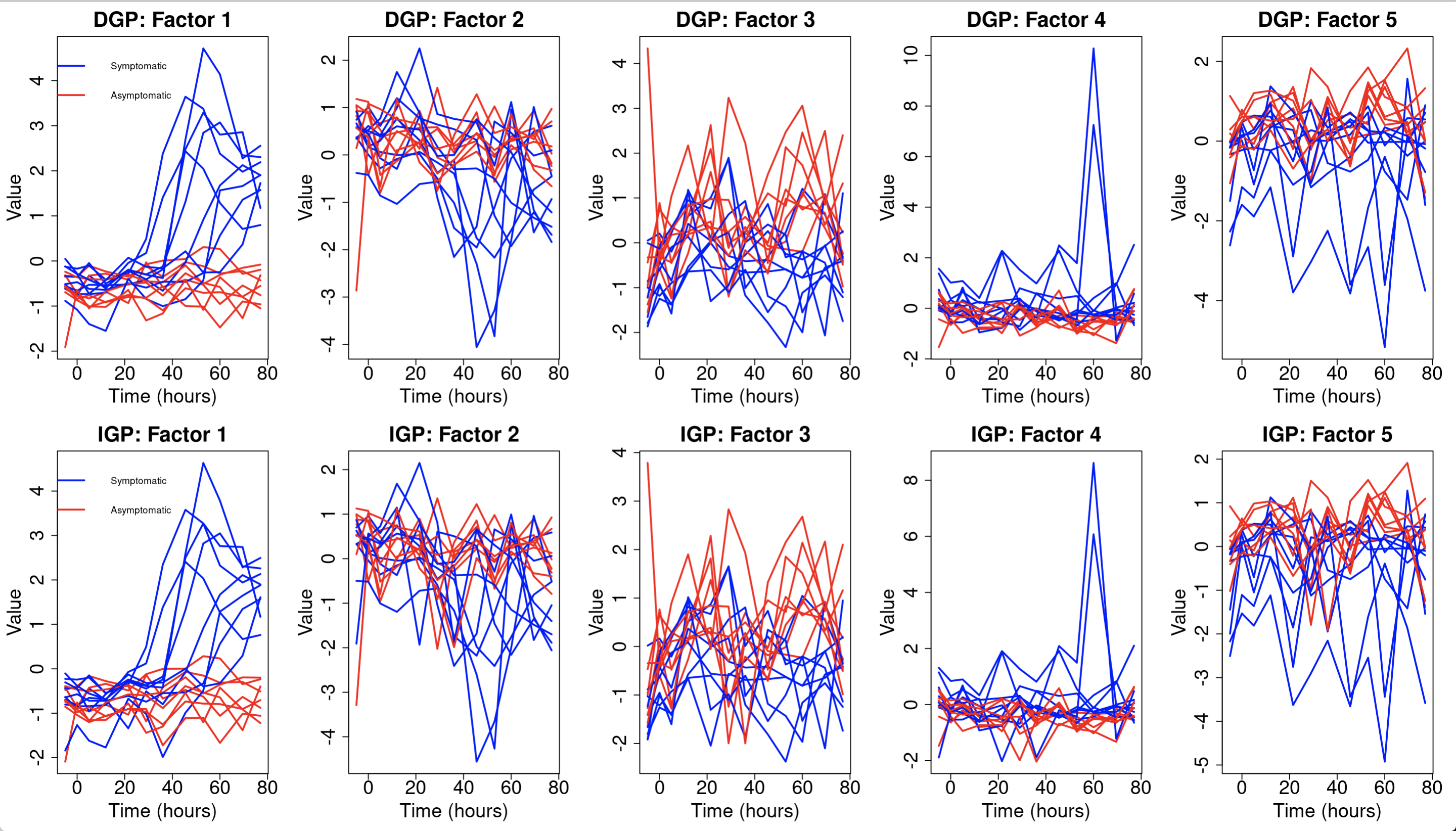}
\caption{Comparison between DGP and IGP on the estimated factor trajectories in the real-data application.}
\label{h3n2_factor_dgp_and_igp}
\end{figure}

\subsection*{D.4 Biological Intepretation}
\begin{figure}[hbt!]
\centering
\includegraphics[width=\textwidth]
{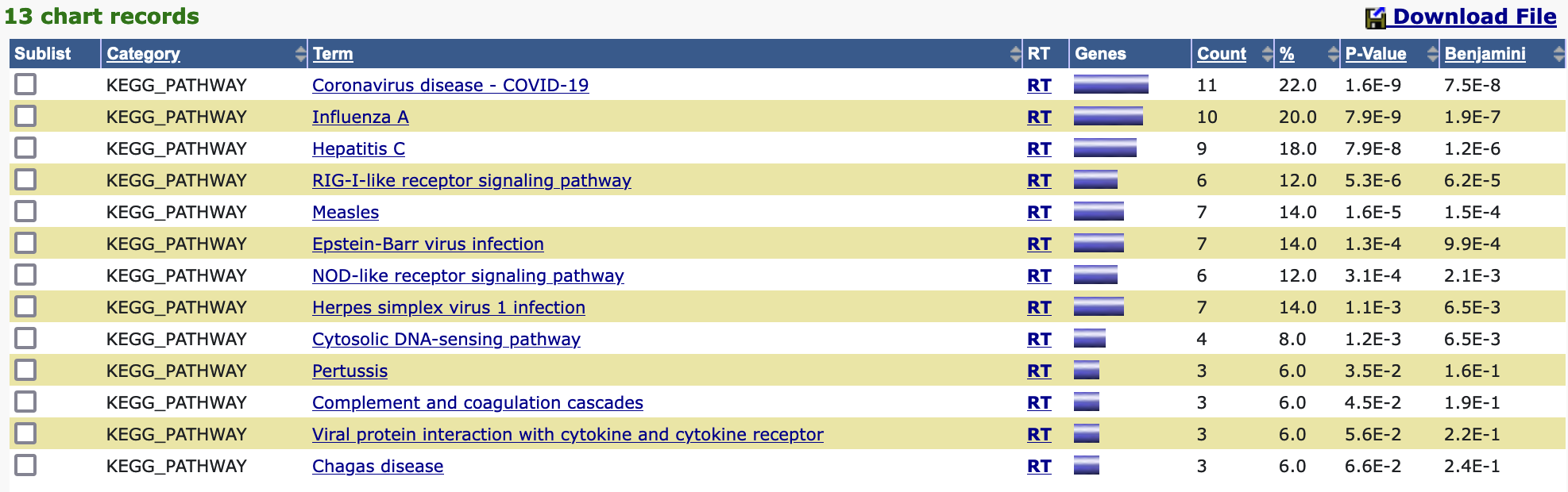}
\caption{Results for genes loaded on factor 1, under the KEGG Pathway analysis of the online bioinformatics platform DAVID.}
\label{factor_1_biology}
\end{figure}

\begin{figure}[htp]
\centering
\includegraphics[width=\textwidth]
{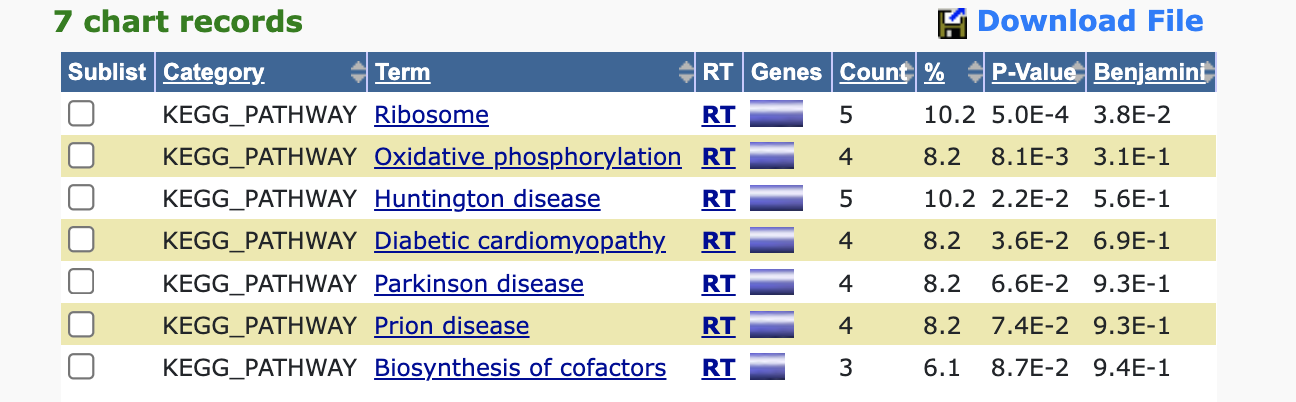}
\caption{Results for genes loaded on factor 3, under the KEGG Pathway analysis of the online bioinformatics platform DAVID.}
\label{factor_3_biology}
\end{figure}

We interpret factor 1 as `innate immune response pathway', because all the significantly enriched pathways are related to immune response (Supplementary Figure~\ref{factor_1_biology}); and interpret factor 3 as the ribosome pathway, because it is the only significantly enriched pathway identified (Supplementary Figure~\ref{factor_3_biology}).

\end{document}